\documentclass[fleqn,10pt]{wlscirep}
\usepackage[utf8]{inputenc}
\usepackage[T1]{fontenc}
\usepackage[numbers,sort&compress]{natbib}
\usepackage{subfigure}
\usepackage{graphicx}
\title{Digital logic from high-efficiency superconducting diodes}

\author[1,*]{Pavan Hosur}
\affil[1]{Affiliation: Department of Physics and Texas Center for Superconductivity, University of Houston, Houston, TX 77204, USA}

\affil[*]{phosur@uh.edu}

\begin{abstract}
Recent advancements in the realizations of superconducting diodes have pushed the diode coefficient $\eta$ towards its theoretical maximum of $\eta=1$. In this work, we describe the construction of logic gates NOT, AND, OR, NAND and NOR using superconducting diodes with $\eta\approx1$ by exploiting their dynamically tunable polarity. We then argue that fundamental theorems suppress $\eta$ in intrinsic superconductors, rendering them likely unsuitable for the proposed devices, and point out that several previous proposals and platforms, remarkably, bypassed this suppression unwittingly. We discuss the realization of the digital logic in one such platform -- Josephson triodes that yielded $\eta\approx1$ -- and argue that phases with spontaneous spatial or magnetic order can overcome some of its drawbacks. Thus, this work provides guiding principles for future platforms and develops the building blocks for superconductors-based digital electronics.
\end{abstract}
\begin{document}

\flushbottom
\maketitle

\thispagestyle{empty}

\section{Introduction}

In recent years, the massive rise in global computational requirements has reinforced the need for low-cost, energy-efficient computing platforms. From this perspective, improvements in cryogenic cooling efficiencies and discoveries of superconductors in more accessible regimes of temperature and pressure inspire superconductors as a promising family of platforms. Moreover, superconductor-based classical computers would integrate naturally with superconducting quantum processors, where cryogenic classical logic has been identified as a potential solution to key challenges in scaling quantum computers, such as latency due to communication between classical and quantum computers and heating \citep{Reilly2019}. So far, the superconducting platform that has been used most for classical computing consists of Josephson arrays, which are inherently hard to scale as they require magnetic fields. Some superconducting transistors have also been realized...

Following the historical development of semiconductors-based digital technology, an alternative natural starting point for its superconducting counterpart is superconducting diodes (SDs), a field that has seen an explosive resurgence of interest in recent years. SDs are defined
by unequal critical current magnitudes $I_{c}^{\pm}$ in opposite
directions and are typically characterized by the diode quality factor
$\eta=\frac{I_{c}^{+}-I_{c}^{-}}{I_{c}^{+}+I_{c}^{-}}\in[0,1]$, whose value has risen rapidly from $\eta\sim 10^{-2}$ in Ref. \citep{Ando:2020td} to, remarkably, $\eta\approx1$
in a Josephson trijunction \citep{Chiles:2023aa}.
The SD effect has been observed in a myriad of platforms \citep{Jiang1994,Ando:2020td,Lyu:2021wg,Du2023,Sundaresh:2023aa,Kealhofer2023,Hou2023,Chen:2024aa,Gupta:2023aa,Baumgartner_2022,Baumgartner:2022wr,Banerjee2023,Pal:2022tm,Kim2024,Turini:2022aa,Bauriedl:2022we,Wu:2022wq,Diez-Merida2021,Golod:2022ta,Chiles:2023aa,Zhang2024,Yun2023,Jeon:2022aa,Lin:2022aa,Anwar:2023aa,Narita:2022tb,Gutfreund:2023aa,Zhao2023,Trahms:2023aa,Yasuda:2019wb,Masuko:2022aa,he2024observationsuperconductingdiodeeffect,Anh:2024aa}
and inspired a wealth of theoretical activity aimed at explaining, enhancing and harnessing SDs for practical purposes \citep{Yuan2022,Daido2022,Daido2022a,Zhang2022,Davydova2022,Chen2024,wang2022symmetry,He_2022,Zhai2022,MIsaki2021,Ilic2022,Scammell_2022,Zinkl2022,He:2023aa,Jiang2022,Kokkeler2022,Debnath2024,Chazono2023,Vodolazov2005,dePicoli2023,Kochan2023,Ikeda2022,Tanaka2022,Wang2022,Haenel2022,Legg2023,Cuozzo2024,Suoto2022,Suoto2024,Cheng2023,Steiner2023,Costa2023,Wei2022,Legg2022,Karabassov2022,Hu2023,Wu2023,Nunchot2024,Daido2023,Cayao2024,Banerjee2024,Hosur2023,chakraborty2024perfectsuperconductingdiodeeffect,Soori2024,Soori_2024}. It relies on a simple
underlying principle to achieve non-reciprocal behavior: broken time reversal, inversion, and any other spatial symmetries that equate ``left'' and ``right''. Crucially, they offer
a tremendous advantage over semiconductor diodes, namely, their polarity
can be dynamically trained with external fields.

In this work, inspired by the trajectory of semiconductors-based
digital technology, we define the building blocks for an analogous path based on SDs. We rely on two key experimental developments: platforms with $\eta\approx1$ and robust control of SD polarity.
We first describe a simple ON/OFF switch with stable memory. 
We build on this design to construct a universal set of
elementary Boolean logic gates, including composite gates where the
output of one level feeds the input of the next. These constructions
form building blocks for complex integrated circuits based on SDs. We then show that in intrinsic SDs, the implicit smoothness in Ginzburg-Landau free energies combined with Bloch's theorem that restricts equilibrium current densities in isolated systems in the thermodynamic limit prevents $\eta\to1$. Thus, natural intrinsic routes to $\eta\approx1$ involve proximity-induced superconductivity and small devices, two approaches that circumvent Bloch's theorem's restrictions in different ways. Alternately, one can use Josephson devices as they are far from the thermodynamic limit and not subject to the above restrictions. Thus, we estimate the required parameters for the Josephson triode where $\eta\approx1$ and tunable polarity have been achieved already \citep{Chiles:2023aa}.

\section{Results}

\begin{figure}
\begin{centering}
\subfigure[]{\begin{centering}
\includegraphics[width=0.25\columnwidth]{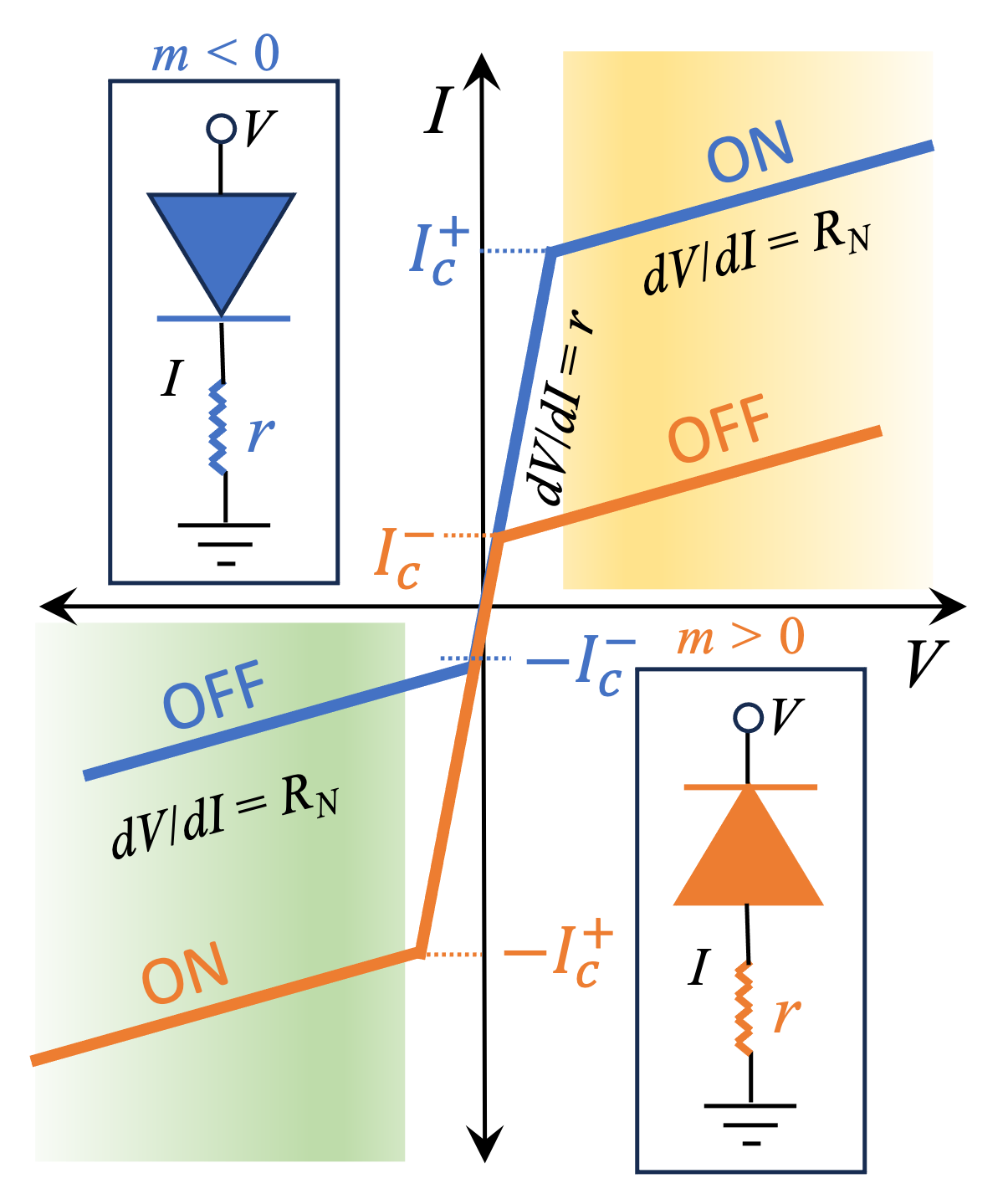}
\par\end{centering}
}\subfigure[]{\begin{centering}
\includegraphics[width=0.25\columnwidth]{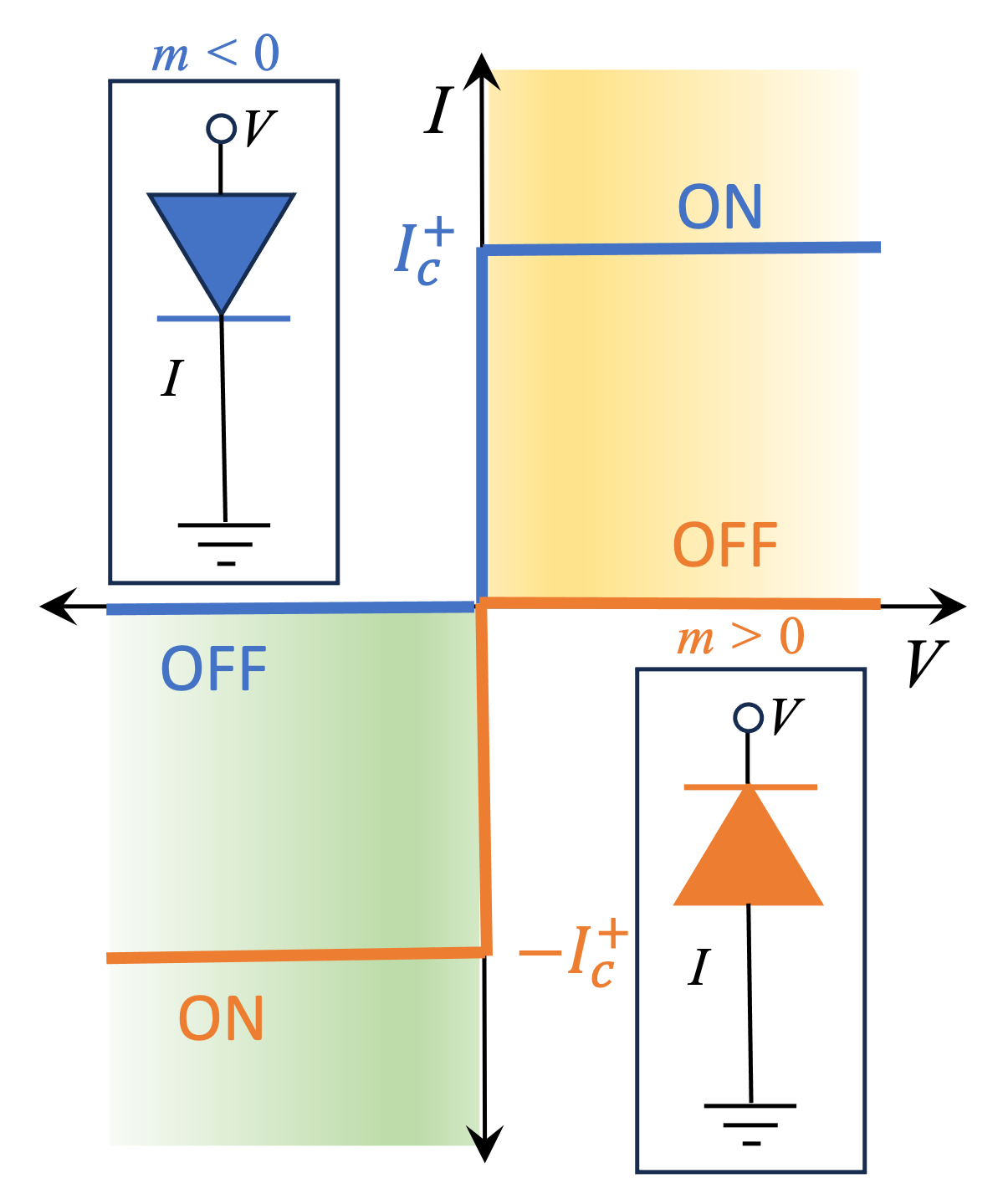}
\par\end{centering}
}

\par\end{centering}
\caption{Schematic $I$-$V$ characteristics of a highly efficient SD trainable by an external Ising field $m$ in non-ideal (a) and ideal (b) limits. The yellow and green shaded regions
offer separate digital switches that can be turned ON/OFF using $m$.
Insets depict the effective SDs for each $\text{sgn}(m)$ with a small
internal resistance $r$. Effective functionality requires $r\ll R_{N}$
and $\epsilon=I_{c}^{-}/I_{c}^{+}\ll1$. The ideal limit is defined by $r=0$, $R_{N}\to\infty$ and $\epsilon=0$.
\label{fig:I-V}}
\end{figure}

Our starting point is an approximation for the current-voltage ($I$-$V$)
characteristics for the SD sketched in Fig. \ref{fig:I-V} in non-ideal
and ideal limits. The polarity of the SD is controlled by an Ising variable $m=\pm$. Denoting the normal state resistance by $R_{N}$ and a small internal resistance of the superconductor due to defects, inhomogeneity etc. by $r\ll R_N$, we assume
\begin{equation}
V_{m}=\begin{cases}
I_{c}^{m}r+(I-I_{c}^{m})R_{N} & I>I_{c}^{m}\\
Ir & -I_{c}^{\bar{m}}<I<I_{c}^{m}\\
-I_{c}^{\overline{m}}r+(I+I_{c}^{\overline{m}})R_{N} & I<-I_{c}^{\bar{m}}
\end{cases}\label{eq:I-V-diode}
\end{equation}
We emphasize three features of this $I$-$V$ relationship that rely on the separation of scales, $I_{c}^{+}\gg I_{c}^{-}$ and $R_{N}\gg r$: (i) in the supertransport regime, $-I_{c}^{-}<|I|<I_{c}^{+}$ near $V=0$, $I$ changes by a large, $O(1)$, amount for small, $O(\epsilon)$, voltage change; (ii) $I$ is nearly independent of $V$ in the normal regimes, $|I|>I_{c}^{+}$. (iii) the $I$-$V$ curve depends on $m$ only through $\text{sgn}(m)$; it has no dependence on $|m|$, which is a fair assumption if the polarity stems from spontaneous order such as ferromagnetism or ferroelectricity whose moment has been saturated by the training field. The precise $I$-$V$ relationship is immaterial if the above salient conditions are satisfied. For instance, nonreciprocal effects in the normal state, such as magnetochiral anisotropy, could be significant when $\epsilon\ll1$ in the SD. Nonetheless, as we show in App. \ref{sec:Normal-state-rectification}, the digital behaviors described in this work are robust in a parametrically large regime.

\subsection{SD-based switch}

To convert the SD into a switch, we exploit the fact that $m$ can
be reversed with an external field. For a fixed, external bias $V$
in the interval $I_{c}^{+}r\lesssim V\ll I_{c}^{+}R_{N}$, the ON/OFF currents are defined as:
\begin{equation}
I\approx\begin{cases}
I_{c}^{+}(\text{ON}) & m<0\\
\epsilon I_{c}^{+}(\text{OFF}) & m>0
\end{cases}\text{ if }I_{c}^{+}r\lesssim V\ll I_{c}^{+}R_{N}\label{eq:I-m-Switch}
\end{equation}
where we have neglected subleading terms in $\epsilon$ and $r/R_{N}$.
Note that using an external field to tune the current between $I_{c}^{+}$
and $I_{c}^{-}$ or reverse its direction is routine for SDs. The
novelties of the above result are the binary ON/OFF definitions and
robustness afforded by the separation of scales between $I_{c}^{+}$
and $I_{c}^{-}$ and between $r$ and $R_{N}$. The upshot is that
the ON/OFF states defined above are insensitive to the precise value of
$V$ as long as it is in the right range. This will allow us to precisely
define Boolean logical states and construct universal gates below.
We assumed $V>0$ in this discussion, but the
situation is identical for $V<0$. Then, the appropriate range of
$V$ is $-I_{c}^{+}R_{N}\ll V\lesssim-I_{c}^{+}r$ while the binary
states are $I\approx-I_{c}^{+}$ (ON) and $I\approx-\epsilon I_{c}^{+}$
(OFF).

\begin{figure}
\begin{centering}
\subfigure[]{\includegraphics[width=0.15\columnwidth]{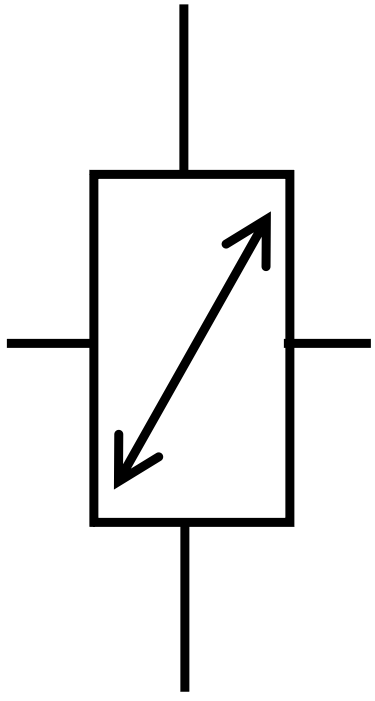}

}~~\subfigure[]{\includegraphics[width=0.23\columnwidth]{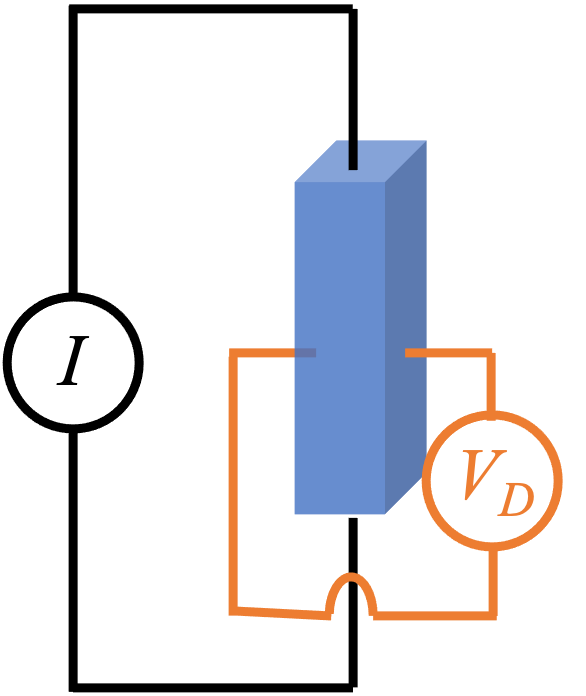}

}
~~\subfigure[]{\includegraphics[width=0.21\columnwidth]{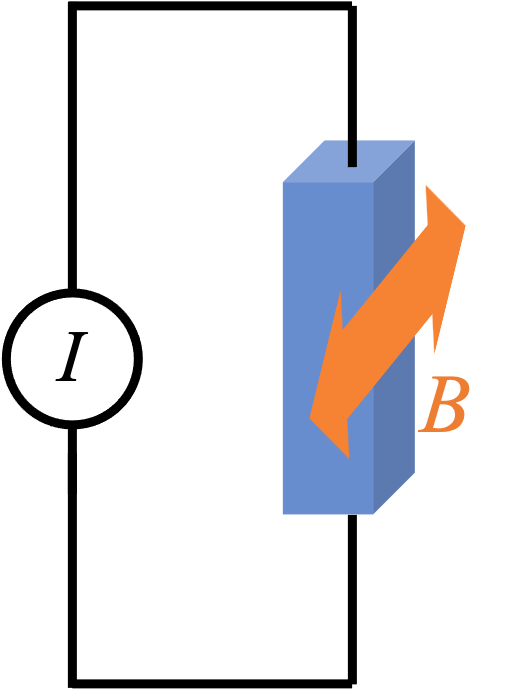}

}
\par\end{centering}
\caption{(a) Circuit symbol for the switch. The field that trains the SD polarity is along the width while transport is along the length. Rightward (leftward) training field yields a SD that superconducts for downward (upward) current and is Ohmic for the opposite current direction. (b,c) Illustration of the switch realized in a quasi-1D structure that exhibits (b) ferroelectricity trainable by the displacement voltage $V_{D}$ or (c) ferromagnetism trainable by a magnetic field $B$. In most realizations, the current, displacement field, and magnetic field are mutually perpendicular. \label{fig:Circuit-symbol}}
\end{figure}

In Fig. \ref{fig:Circuit-symbol}, we introduce a circuit symbol for
an ideal SD-based switch, which must be read as follows. The 
field that trains $m$ -- such as an out-of-plane (in-plane) displacement (magnetic) field in 2D systems with out-of-plane polarization (in-plane magnetization) -- is applied along the shorter side of the rectangle while current flows along the longer side. The diagonal double-headed arrow depicts the biasing conditions. Specifically, a rightward or positive (leftward or negative) $m$ yields a SD whose polarity is upward (downward).

\subsection{Single-input NOT gate}
 
To construct Boolean logic gates using the SD, we first introduce
a load resistance $R$ that satisfies $r\ll R\ll R_{N}$. Then, we
define logical states as given in Table \ref{tab:Logical-definitions}.
Intuitively, large and small voltages -- $O\left(1\right)$ and $O(\epsilon,R/R_{N})$,
respectively -- are defined as logical 1 and 0. For the Ising variable $m$, we arbitrarily choose $m<0$ ($m>0$) as logical 0 (1).

\begin{table}
\begin{centering}
\begin{tabular}{|c|c|c|c|}
\hline 
Input Ising variable & Input physical variable & Output voltage & Logical value\tabularnewline
\hline 
\hline 
$m<0$ & $V_D\sim I_{c}^{+}R\times O(\epsilon,R/R_{N})$ or $B<0$ & $\sim I_{c}^{+}R\times O(\epsilon,R/R_{N})$ & 0\tabularnewline
\hline 
$m>0$ & $V_D\sim I_c^+R$ or $B>0$ & $\sim I_{c}^{+}R$ & 1\tabularnewline
\hline 
\end{tabular}
\par\end{centering}
\caption{Input and output variables and their corresponding to the logical states. Input can be any binary Ising field, denoted $m$ here, while outputs are voltages. To develop circuits using SDs, the Ising input must be mapped to the binary voltage values that define the logical output. The appropriate regime for digital logical operation is $r\ll R\ll R_{N}$ and $\epsilon\ll1$.
\label{tab:Logical-definitions}}
\end{table}
Secondly, to build digital circuits where the outputs at one layer
of gates serve as inputs to the next layer, $\text{sgn}(m)$ must
ultimately be electrically trainable. If $m$ denotes magnetization, a magneto-electric or an electromagnet would be additionally needed to convert between electric and magnetic signals. In either case, we require that $m$ is saturated by a voltage of $O(I_{c}^{+}R)$; specifically, by $\pm I_{c}^{+}R/2$ as we will see shortly. We will assume $I_{c}^{+}$ to be large enough that $\pm I_{c}^{+}R/2$ can train $\text{sgn}(m)$ reliably. Note, $R$ cannot be made too large as it is limited by $R_{N}$ and Ohmic heating must be avoided as the devices are based on a SDs.

\begin{figure}
\begin{centering}

\subfigure[]{\includegraphics[width=0.25\columnwidth]{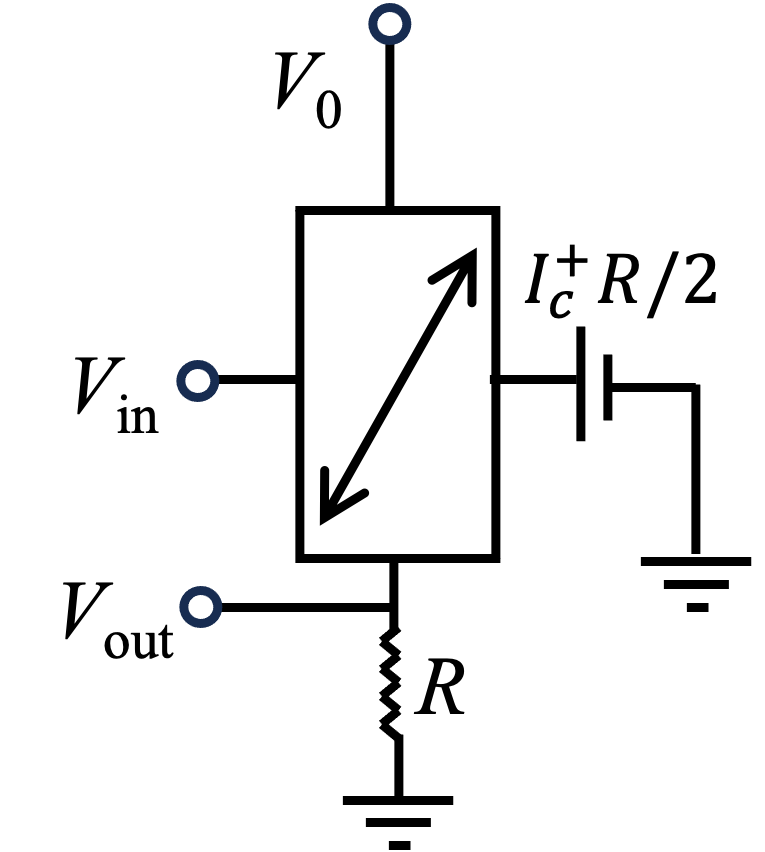}

}\subfigure[]{\includegraphics[width=0.25\columnwidth]{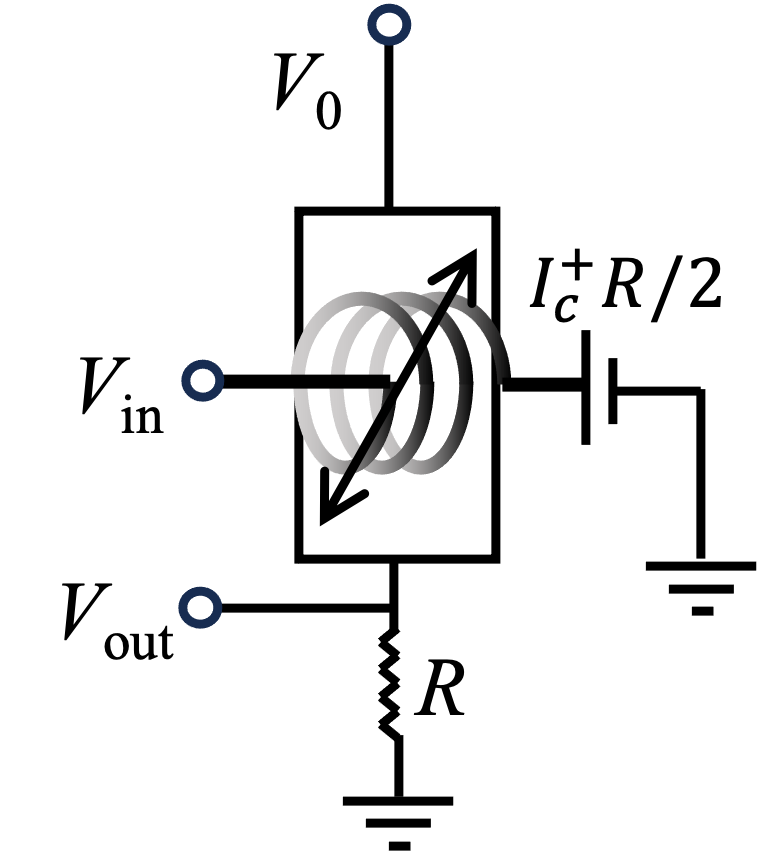}

}\subfigure[]{\includegraphics[width=0.155\columnwidth]{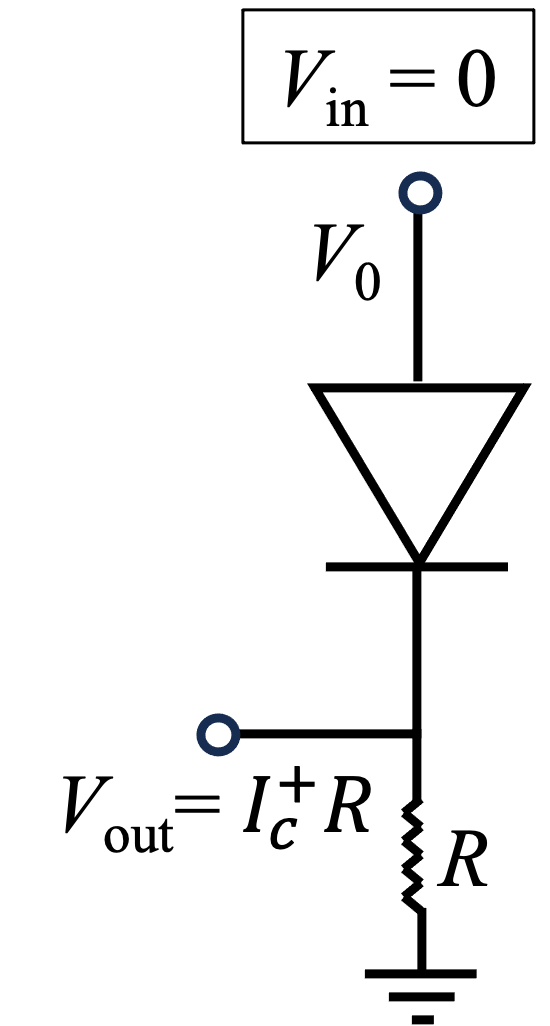}

}\subfigure[]{\includegraphics[width=0.15\columnwidth]{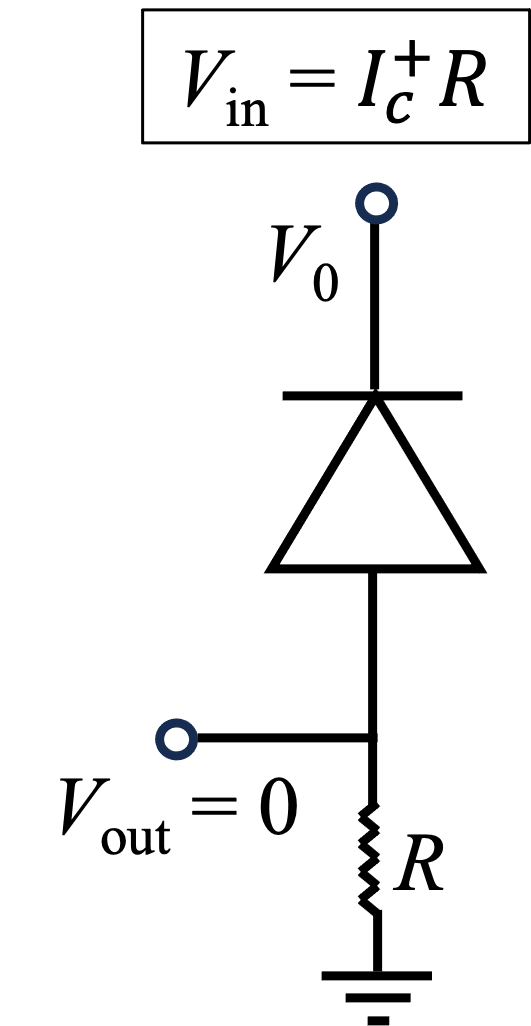}

}
\par\end{centering}
\caption{The NOT gate constructed from a high-efficiency SD whose polarity  controlled by (a) an electric or (b) a magnetic field, and its effective diode depictions for input 0 (c) and input 1 (d). The double-headed arrow indicates the relationship between the diode polarity and the training field as in Fig. \ref{fig:Circuit-symbol}. Digital operation requires $I_{c}^{+}R_{N}\gg V_{0}>I_{c}^{+}R$.\label{fig:NOT}}
\end{figure}

The definitions above immediately yield the simplest logic gate, the
single bit NOT gate, shown in Fig. \ref{fig:NOT}. In the presence
of a bias voltage $V_{0}$ as shown, the input and output voltages in Fig. \ref{fig:NOT}(a) can be straightforwardly be shown to satisfy the following exact relations assuming Eq. (\ref{eq:I-V-diode}) (see App. \ref{sec:Voltage-relations} for details): 
\begin{equation}
V_{\text{out}}=\begin{cases}
\frac{V_{0}+I_{c}^{+}(R_{N}-r)}{R+R_{N}}R & V_{\text{in}}<I_{c}^{+}R/2\\
\frac{V_{0}+I_{c}^{-}(R_{N}-r)}{R+R_{N}}R & V_{\text{in}}>I_{c}^{+}R/2
\end{cases}\label{eq:NOT-exact}
\end{equation}
To first order in $1/R_{N}$ and $\epsilon$, these expressions simplify
to
\begin{equation}
V_{\text{out}}\approx\begin{cases}
I_{c}^{+}R\left(1+\frac{V_{0}/I_{c}^{+}-R-r}{R_{N}}\right)\equiv1 & V_{\text{in}}\equiv0\\
\epsilon I_{c}^{+}R+V_{0}\frac{R}{R_{N}}\equiv0 & V_{\text{in}}\equiv1
\end{cases}
\end{equation}
where $V_{\text{in}}\equiv0$ and 1 correspond to the ranges $V_{\text{in}}<I_{c}^{+}R/2$
and $V_{\text{in}}>I_{c}^{+}R/2$ and denote the appropriate logical
values per the definitions in Table \ref{tab:Logical-definitions}.
Thus, setting $V_{0}$ in the range $I_{c}^{+}(R+r)\lesssim V_{0}\ll I_{c}^{+}R_{N}$
turns the switch into a NOT gate whose 1 (0) output corresponds to
the ON (OFF) states of the switch. In App. \ref{sec:Normal-state-rectification},
we show that including nonreciprocal effects in the normal state,
quantified by a coefficient $\alpha$ with units of inverse current, merely raises the lower limit on $V_{0}$ to $\sim I_{c}^{+}(R+r)/(1-I_{c}^{+}|\alpha|)$. %

Intuitively, $V_{\text{in}}\approx0$ makes the SD forward-biased
and drives a large downward current $\sim I_{c}^{+}$, ensuring a
large $V_{\text{out}}\approx I_{c}^{+}R$ which corresponds to a logical
1. If $V_{\text{in}}\approx I_{c}^{+}R$ on the other hand, the SD
is reverse-biased, the downward current is $\sim\epsilon I_{c}^{-}+V_{0}/R_{N}$
and $V_{\text{out}}\approx(\epsilon I_{c}^{+}+V_{0}/R_{N})R$ is correspondingly
small, representing logical 0.

\subsection{Two-input gates}

Unlike semiconductor transistors and diodes, the current through the
SD in the operating regime is fixed at $I_{c}^{+}$ or $0$ to zeroth
order in $1/R_{N}$ and $\epsilon$. This restricts the allowed combinations
of SDs and prevents, for instance, creating an OR gate by connecting
inputs in parallel as is standard for semiconductor transistor-based
gates. We now illustrate the design for the two-input gates NOR, AND,
OR and NAND. Since all digital circuits can in principle be constructed
from NOR or NAND, this section enables the leap from SD-based binary
logic to digital computing in this platform.

\begin{figure}
\begin{centering}
    
\subfigure[]{\includegraphics[width=0.2\columnwidth]{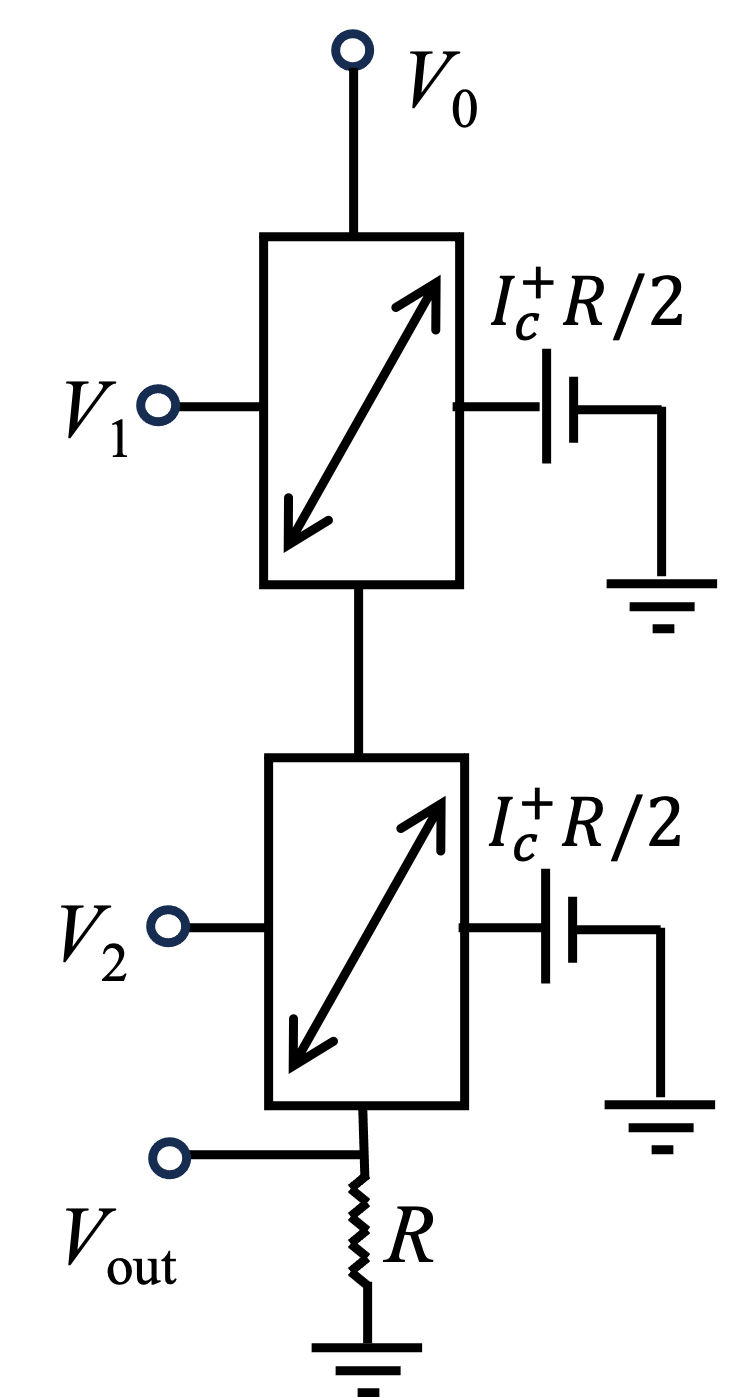}

}\subfigure[]{\includegraphics[width=0.2\columnwidth]{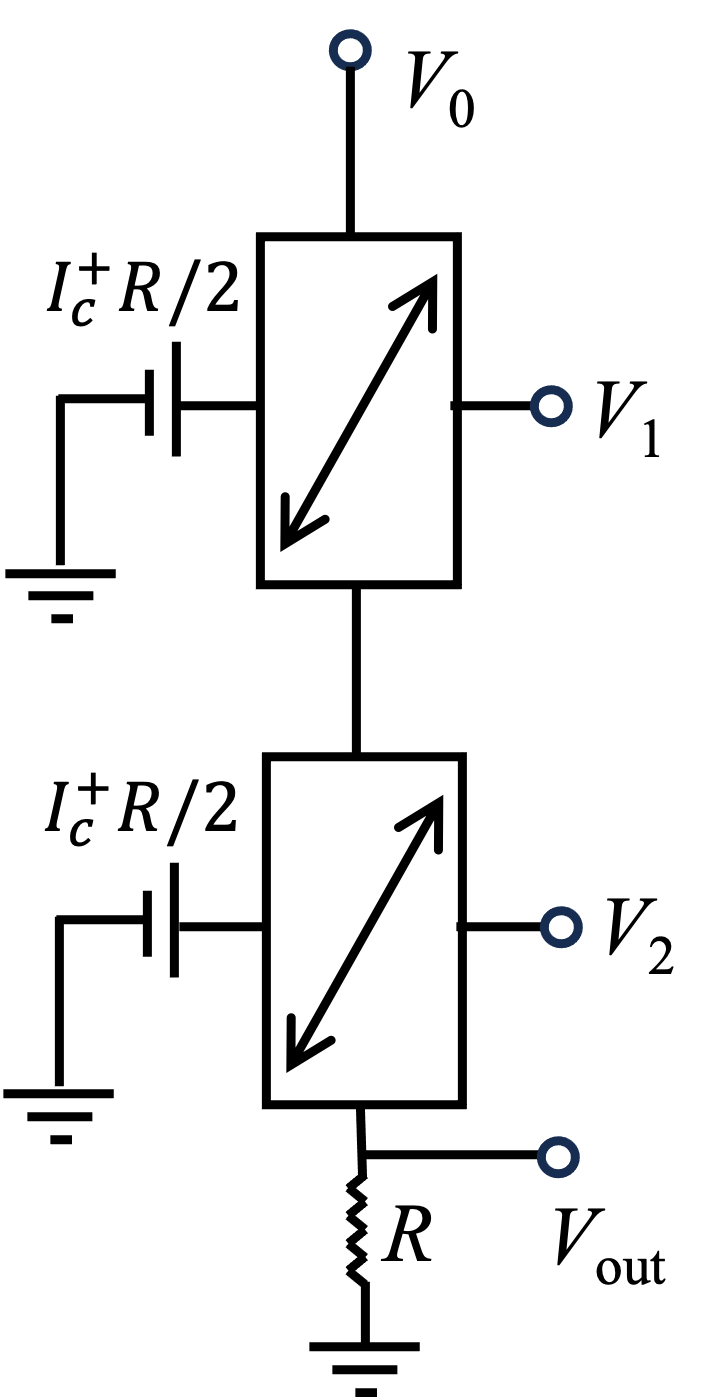}

}\subfigure[]{\includegraphics[width=0.27\columnwidth]{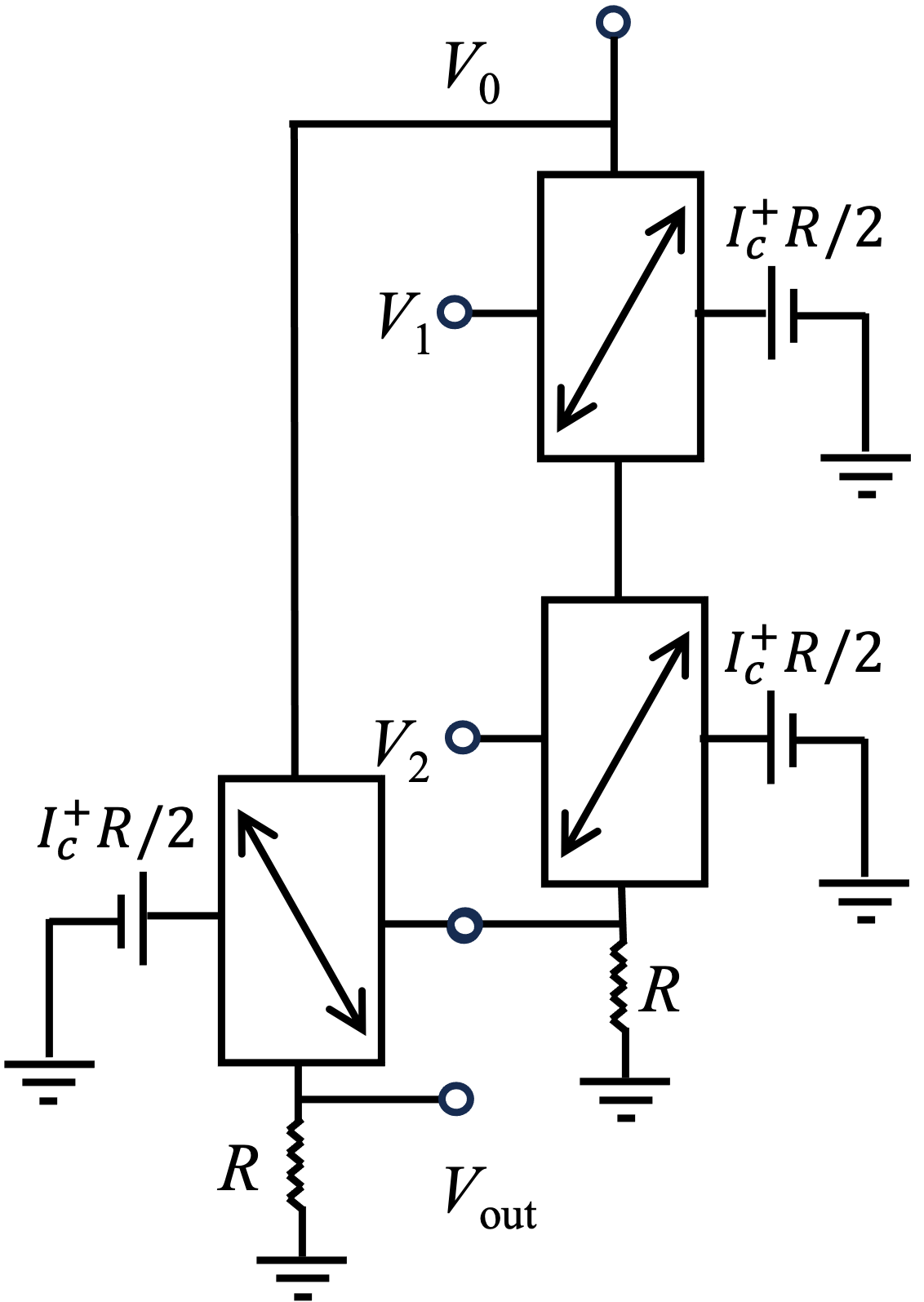}

}\subfigure[]{\includegraphics[width=0.27\columnwidth]{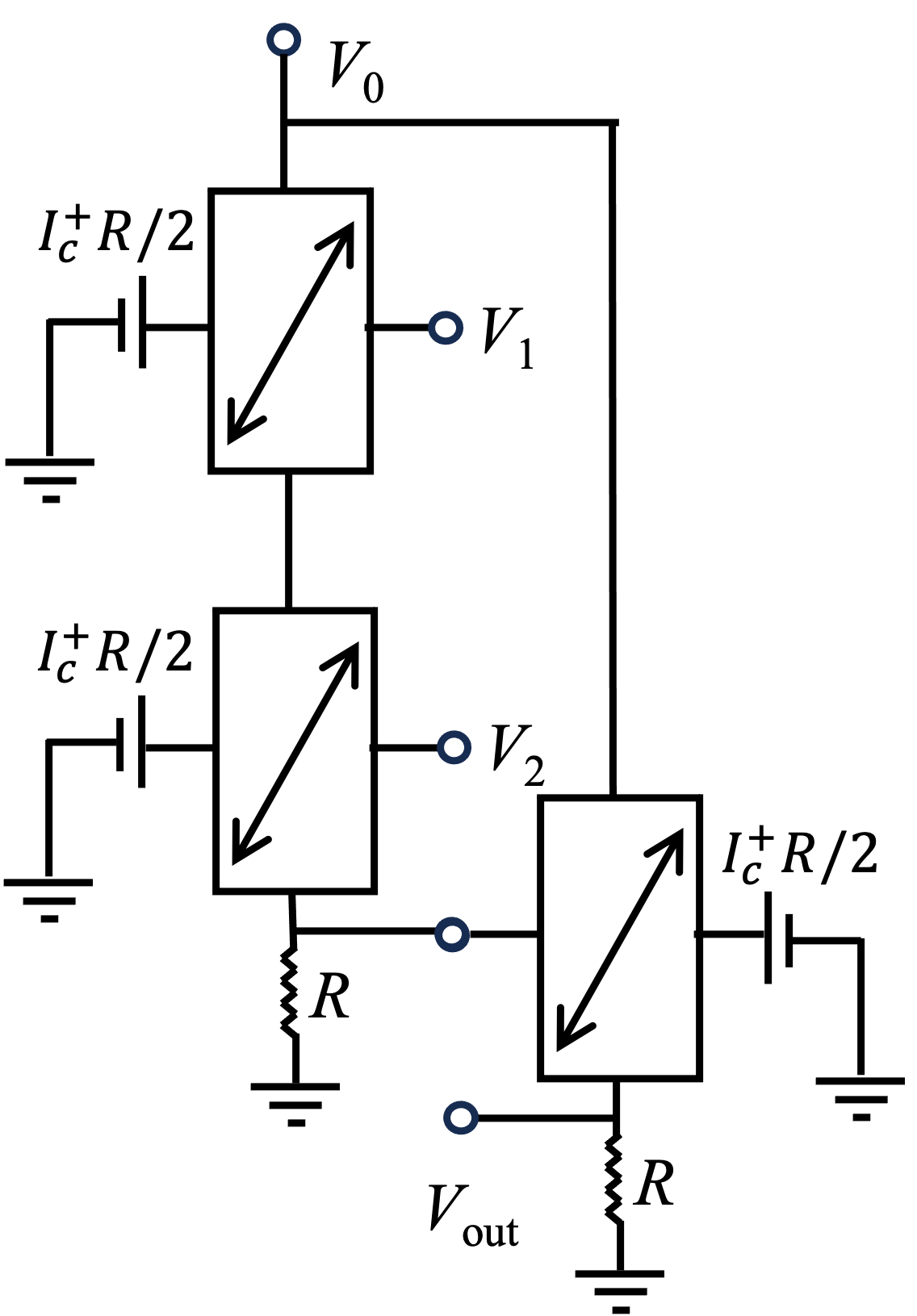}

}
\par\end{centering}
\caption{Two-input gates. (a) NOR (b) AND (c) OR (d) NAND. Negating the inputs of NOR (a) and NAND (c) with a NOT yield AND (b) and NAND (d), respectively, and vice-versa. Appending a NOT at the output of NOR (a) and AND (b) yields OR (c) and NAND (d), respectively.
respectively. \label{fig:Two-bit gates}}
\end{figure}

A similar analysis as that performed for NOT allows the construction of the NOR
and the AND gates, see Fig. \ref{fig:Two-bit gates}(a,b). For brevity, we have only shown constructions based on electrically controllable SDs. Now, the inputs and outputs are related as
\begin{equation}
V_{\text{out}}^{\text{NOR}}\approx\begin{cases}
I_{c}^{+}R\left(1+\frac{V_{0}/I_{c}^{+}-R-2r}{2R_{N}}\right) & V_{\text{in}}\equiv0,0\\
\epsilon I_{c}^{+}R+V_{0}\frac{R}{R_{N}} & V_{\text{in}}\equiv0,1\\
\epsilon I_{c}^{+}R+V_{0}\frac{R}{R_{N}} & V_{\text{in}}\equiv1,0\\
\epsilon I_{c}^{+}R+V_{0}\frac{R}{2R_{N}} & V_{\text{in}}\equiv1,1
\end{cases}
\end{equation}
where $V_{\text{in}}\equiv0,0$ denotes $V_{1}<I_{c}^{+}R/2$, $V_{2}<I_{c}^{+}R/2$,
etc. For the AND gate, the corresponding relations follow by reversing
all the inequalities for $V_{\text{in}}$, since AND is equivalent to NOR with all the inputs negated:
\begin{equation}
V_{\text{out}}^{\text{AND}}\approx\begin{cases}
I_{c}^{+}R\left(1+\frac{V_{0}/2I_{c}^{+}-R/2-r}{R_{N}}\right) & V_{\text{in}}\equiv1,1\\
\epsilon I_{c}^{+}R+V_{0}\frac{R}{R_{N}} & V_{\text{in}}\equiv1,0\\
\epsilon I_{c}^{+}R+V_{0}\frac{R}{R_{N}} & V_{\text{in}}\equiv0,1\\
\epsilon I_{c}^{+}R+V_{0}\frac{R}{2R_{N}} & V_{\text{in}}\equiv0,0
\end{cases}
\end{equation}
Negating the outputs of NOR and AND yield OR and NAND, respectively, as shown in Figs. \ref{fig:Two-bit gates}(c,d). In these cases,
negation requires explicitly appending a NOT at the output. This is
safe as long as the current through the transverse terminals is negligible or absent. Since NOR and NAND are universal gates, any digital circuit can be built in principle using SDs. Detailed analysis of all the gates is given in App. \ref{sec:Voltage-relations}.

\section{Discussion}\label{sec:Discussion}

\subsection{Barrier to $\epsilon\ll1$}

Our results rely fundamentally on the condition $\epsilon=I_c^-/I_c^+\ll1$, i.e., a low OFF/ON ratio, resulting in a small leakage current. However, we now show that basic principles hinder a small $\epsilon$ in intrinsic SDs.

Intrinsic SDs are typically described in theory by a Cooper pair momentum ($q$) dependent condensation energy density $F[q]$. Superconductivity exists only if $F[q]<0$, and supercurrent density is given by $j(q)=2e\partial_q F[q]$ whose largest positive and negative values in opposite directions are the critical current densities $j_c^\pm$. This approach is convenient in theory as it follows directly from a Bogoliubov-deGennes Hamiltonian that includes $q$-dependent pairing. However, it overlooks the fact that experiments invariably control $j$, not $q$. 

Instead, consider the Gibbs condensation energy
\begin{equation}
    G[j] = F[q(j)] - \frac{1}{2e}jq(j)
\end{equation}
Again, superconductivity exists only when $G[j]<0$. A fundamental theorem by Bloch forbids equilibrium current densities in arbitrary isolated quantum systems \citep{Watanabe:2019aa}. This mandates a stable equilibrium state at $j=0$, corresponding to the conditions $G'[0]=0$, $G''[0]>0$. Moreover, $G[0]<0$ for this state to be superconducting while $j_c^\pm$ are magnitudes of the smallest positive and negative roots of $G[j]$, which implies $G[j_c^\pm]=0$. The upshot of these conditions is that a large $\eta$ or small $\epsilon$, defined by $j_c^\pm\gg j_c^-$, requires $G[j]$ to change sharply around $j=0$. In particular, the limit $\eta=1$ requires $j_c^-=0$ and $j_c^+\neq0$, which is inconsistent with $G[0]<0$ unless $G[j]$ is non-analytic at $j=0$, implying a phase transition distinct from the superconducting transition at $j=0$. Indeed, Ref. \citep{Yuan2022} and \citep{chakraborty2024perfectsuperconductingdiodeeffect} showed that $\eta=1$ at a tricritical point in superconducting Rashba metals and along a critical line in superconducting altermagnets, respectively. However, these proposals require fine-tuning to criticality and motivate a search for alternatives.

\subsection{Experimental platforms with large $\eta$}

One such route is through the proximity effect, which offers the added advantage of providing a natural substrate for patterning the desired circuit. Refs. \citep{Hosur2023,yuan2023surfacesupercurrentdiodeeffect} exploited this effect to propose routes to $\eta=1$ on superconductor surfaces without fine-tuning. On the experimental front, nanowires of superconducting $\beta$-Sn were recently patterned on the surface of the Dirac semimetal $\alpha$-Sn \citep{Anh:2024aa}. The wires showed a large SD effect with a relatively $\eta\sim25\%$ trainable by a longitudinal magnetic field. Moreover, the effect was argued to arise from the interface between the nanowires and the semimetal. In this setup, Bloch's theorem constrains the whole system, not the nanowire or the interface separately. Thus, $G^\prime_\text{wire}[0]$ can be non-zero, allowing $j_c^\pm\gg j_c^-$ without $G_\text{wire}[j]$ relinquishing its analyticity. 

Another route to a large $\eta$ is via relatively small devices. This is because Bloch's theorem bounds the equilibrium current density by the inverse longitudinal dimension of the system \citep{Watanabe:2019aa}, so short systems can carry non-negligible equilibrium currents. This idea likely helped reach $\eta\sim65\%$ in heterostructures of ferromagnetic EuS and superconducting V deposited on Pt \citep{Hou2023}. Here, asymmetries in the sample edges were key to obtaining a large SD effect, suggesting the importance of finite system size. This platform was further developed into multi-SD circuits for full wave rectification with $\sim43\%$ efficiency and ac-to-dc conversion up to 40kHz \citep{Ingla-Aynes2024}.

Among existing platforms, arguably a strong candidate for realizing the proposals in this work is the Josephson triode made of MoRe-graphene-MoRe Josephson junctions \citep{Chiles:2023aa}. Here, the microscopic physics is likely highly quantized and beyond the thermodynamic considerations of Ginzburg-Landau theory, such as the smoothness of free energy. In this device, a diode effect between two terminals was controlled by the current through a third terminal. Crucially, a representative device achieved $\eta\approx1$ with $I_c^+\approx0.30$ $\mu\text{A}$ and training current $I_t\approx0.16$ $\mu\text{A}$, while $r$ was negligible compared to $R_N\approx 200$ $\Omega$. For these parameters, the designs we propose require $R\ll 200$ $\Omega$ and $V_0\gtrsim I_c^+R\sim 60$ $\mu\text{V}$, which are easily attainable. On the other hand, the Josephson triode has important drawbacks that hinder its scalability for digital circuits. In particular, $I_t$ is comparable to the ON current $I_c^+$ and can lead to significant stray currents, unlike semiconductor transistor-based circuits where the bias current is negligible compared to the ON current. In addition, $I_c^+$ varies smoothly with $I_t$, making it vulnerable to noise in $I_t$. 

For immunity against such noise, one would like the training field to couple to and saturate the order parameter of a spontaneously broken symmetry. In particular, a ferroelectric superconductor with additional time-reversal symmetry breaking would form an ideal platform, since it would allow electric control of SD polarity and the voltage output of one circuit layer can directly feed the input of the next. Electrostatic control of SD was recently achieved in Josephson diodes made of epitaxial Al-InAs junctions \citep{shin2024electriccontrolpolarityspinorbit}, but the control knob was a gate voltage on which $I_c^+$ depends smoothly and will presumably inherit noise from as in the Josephson triode. On the other hand, Ref. \citep{Zhai2022} proposed a SD based on a ferroelectric metal, bilayer NbSe$_{2}$ intercalated with Cu. Such metals were predicted in 1965 \citep{Anderson1965} but had been elusive until the recent discoveries of ferroelectricity in WTe$_{2}$ \citep{Fei:2018aa,Barrera:2021tb} and bilayer graphene/boron nitride moire systems \citep{Zheng:2020ui}. Recently, coupling between ferroelectricity and superconductivity was observed in bilayer MoTe$_{2}$ \citep{Jindal:2023uh} while simultaneous ferromagnetism and electric polarity were seen for the first time in a metal, (Fe$_{0.5}$Co$_{0.5}$)$_{5}$GeTe$_{2}$ \citep{ZhangFCGT}, raising prospects of intrinsic or proximity-induced superconductivity in it to create a SD that is robustly trainable by both electric and magnetic fields. Along with the logic and designs described in this work, the above developments lay the groundwork and motivate a search for platforms for superconducting digital electronics.

\section{Conclusions}

We have tackled two key problems on the road to superconducting digital technology. First, we constructed circuits for logic gates based on high-efficiency SDs using only resistors and voltage sources. The key property of SDs that we exploit is their dynamically reversible polarity. We explicitly constructed NOT, AND, OR, NAND and NOR gates, thus providing several sets of universal building blocks for digital computing. Secondly, we discussed how fundamental principles reduce SD efficiency in intrinsic systems, thereby narrowing the search space of platforms where high efficiency is possible, and point out how an implicit circumvention of these principles led to high efficiency in some existing platforms. The proposed digital properties should be achievable in MoRe-graphene-MoRe Josephson triodes \citep{Chiles:2023aa}; while scalable circuits will likely need trainable broken symmetry phases such as ferroelectrics and ferromagnets. In sum, this work paves the way for digital technology using superconducting diodes.

\section*{Acknowledgements}

We thank Ramamoorthy Ramesh, Gururaj Naik, Ashvin Vishwanath, Nandini Trivedi, and Pedram Roushan for useful discussions. We are grateful to the Department of Physics and Astronomy, Rice University, and the Max Planck Institute for the Physics of Complex Systems for their hospitality during different parts of this project. This research was supported by the Department of Energy Basic Energy Sciences grant no. DE-SC0022264.

\appendix

\part*{Appendix}

\section{Effect of normal state rectification}\label{sec:Normal-state-rectification}
Unless time-reversal, inversion, and appropriate spatial symmetries
are broken only in the superconducting state, systems that exhibit
a SD effect also perform rectification in the normal state. In other
words, their normal state resistance is current dependent which in the simplest case can be written as:
\begin{equation}
R_{N}(I)\equiv\frac{dV}{dI}=R_{0}(1+\alpha I)
\end{equation}
A well-known example of such an effect is magnetochiral anisotropy
where $\alpha I\equiv(\boldsymbol{\gamma}\times\mathbf{B})\cdot\mathbf{I}$,
$\mathbf{B}$ being an external magnetic field and $\boldsymbol{\gamma}$
known as the magnetochiral anisotropy coefficient. Normal state rectification
relies on the same symmetries as the SD effect, so it is conceivable
that it would be significant in systems where the latter is highly
efficient, i.e., $\epsilon\ll1$. On the other hand, it is also known
that the ratio of $\gamma$'s in the fluctuating-superconductor and normal regimes
scales as \citep{Wakatsuki2017}
\begin{equation}
\frac{\gamma_{SC}}{\gamma_{N}}\sim\left(\frac{E_{F}}{k_{B}T_{c}}\right)^{3}
\end{equation}
where $E_F$ is the normal state Fermi energy and $T_c$ is the superconducting transition temperature. As a result, it can grow by several orders of magnitude as superconductivity onsets \citep{Yasuda:2019wb,Masuko:2022aa}.
Thus, a small value of $|\alpha|\equiv|\gamma B|$, a normal
state property, is physically consistent a large diode effect, $\eta\sim1$
or $\epsilon\ll1$. 

We now explicitly re-derive the characteristics
of the NOT gate and prove that its digital behavior is robust as long
as $|\alpha|\ll1/I_{c}^{+}$. The basic $I$-$V$ characteristics of the SD, including the resistance non-linearity, read:

\begin{equation}
V_{m}=\begin{cases}
I_{c}^{m}r+(I-I_{c}^{m})R_{N}(I) & I>I_{c}^{m}\\
Ir & -I_{c}^{\bar{m}}<I<I_{c}^{m}\\
-I_{c}^{\overline{m}}r+(I+I_{c}^{\overline{m}})R_{N}(I) & I<-I_{c}^{\bar{m}}
\end{cases}\label{eq:I-V-diode-1}
\end{equation}
while the NOT gate characteristics become
\begin{align}
V_{0}=IR+I_{c}^{+}r+(I-I_{c}^{+})R_{N}(I) & \text{if }V_{\text{in}}<I_{c}^{+}R/2\text{ and }V_{0}-IR>I_{c}^{+}r\\
V_{0}=IR+I_{c}^{-}r+(I-I_{c}^{-})R_{N}(I) & \text{if }V_{\text{in}}>I_{c}^{+}R/2\text{ and }V_{0}-IR>I_{c}^{-}r
\end{align}
Solving for $I$,
\begin{align}
I & =\frac{-\left[R+R_{0}(1-I_{c}^{\pm}\alpha)\right]+\sqrt{\left[R+R_{0}(1-I_{c}^{\pm}\alpha)\right]^{2}+4R_{0}\alpha\left[I_{c}^{\pm}(R_{0}-r)+V_{0}\right]}}{2R_{0}\alpha}
\end{align}
for $V_{0}-IR>I_{c}^{\pm}r$, %
where we have selected the root of the quadratic equation that is
smoothly connected to the solution in the $\alpha\to0$ limit. For
$R_{0}\gg R$ and $|\alpha|I_{c}^{\pm}\ll1$, this simplifies to
\begin{equation}
I\approx\frac{I_{c}^{\pm}R_{0}+V_{0}}{R_{0}(1-I_{c}^{\pm}\alpha)}
\end{equation}
Self-consistency requires $V_{0}-IR>I_{c}^{\pm}r$ or
\begin{equation}
V_{0}\gtrsim\frac{I_{c}^{\pm}R}{1-I_{c}^{\pm}\alpha}+I_{c}^{\pm}r
\end{equation}
Thus, a larger bias voltage, $V_{0}\gtrsim\frac{I_{c}^{+}R}{1-|\alpha|I_{c}^{+}}>I_{c}^{+}R$
ensures that the digital operation of the NOT gate as described in
this paper. The analysis above also establishes that the normal
state non-linearity merely demands an enhanced $V_{0}$ without qualitatively changing the physics. Therefore, other logic gate operations
should also be similarly robust.

\section{Voltage relations for various gates\label{sec:Voltage-relations}}

In this section, we supply the details that lead to the input-output
voltage relations stated in the main text. The derivations are based
Kirchoff's laws; the main departure from introductory physics is that
Ohm's law is replaced by the SD $I$-$V$ relation given in Eq. \ref{eq:I-V-diode}
of the main text.

Explicitly, for the single input NOT gate in Fig. \ref{fig:NOT},
we have
\begin{align}
V_{0}=IR+I_{c}^{+}r+(I-I_{c}^{+})R_{N} & \text{if }V_{\text{in}}<I_{c}^{+}R/2\text{ and }V_{0}-IR>I_{c}^{+}r\\
V_{0}=IR+I_{c}^{-}r+(I-I_{c}^{-})R_{N} & \text{if }V_{\text{in}}>I_{c}^{+}R/2\text{ and }V_{0}-IR>I_{c}^{-}r
\end{align}
where $I$ is the downward current. Above, the condition
on $V_{\text{in}}$ implements the correct polarity of the SD while
that on $V_{0}-IR$ ensures self-consistency that the SD is in the
assumed (normal) regime. With $V_{\text{out}}=IR$, this immediately
yields Eq. (\ref{eq:NOT-exact}) in the main text.

\begin{figure*}
\begin{centering}
\subfigure[]{\includegraphics[width=0.25\columnwidth]{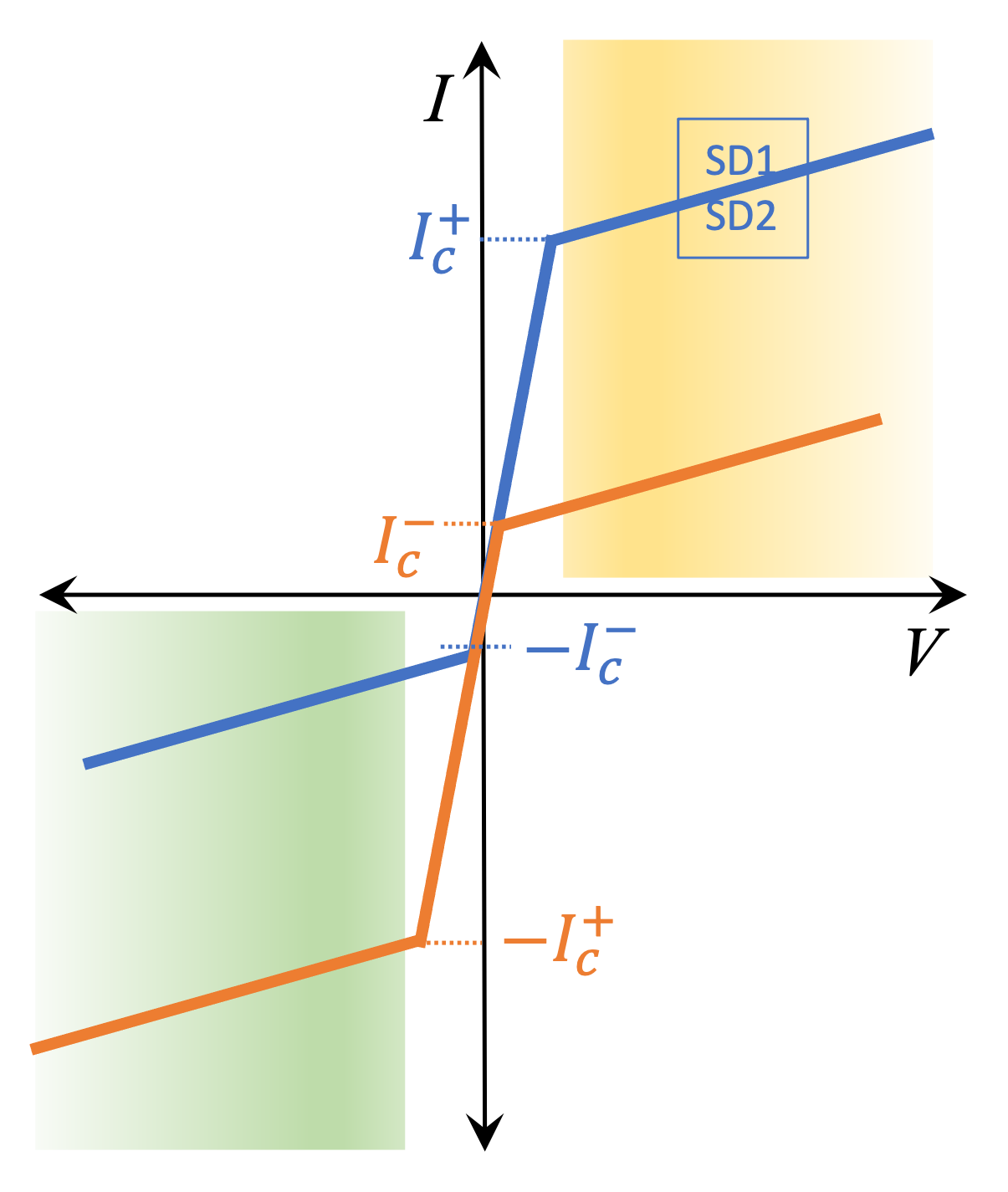}
}\subfigure[]{\includegraphics[width=0.25\columnwidth]{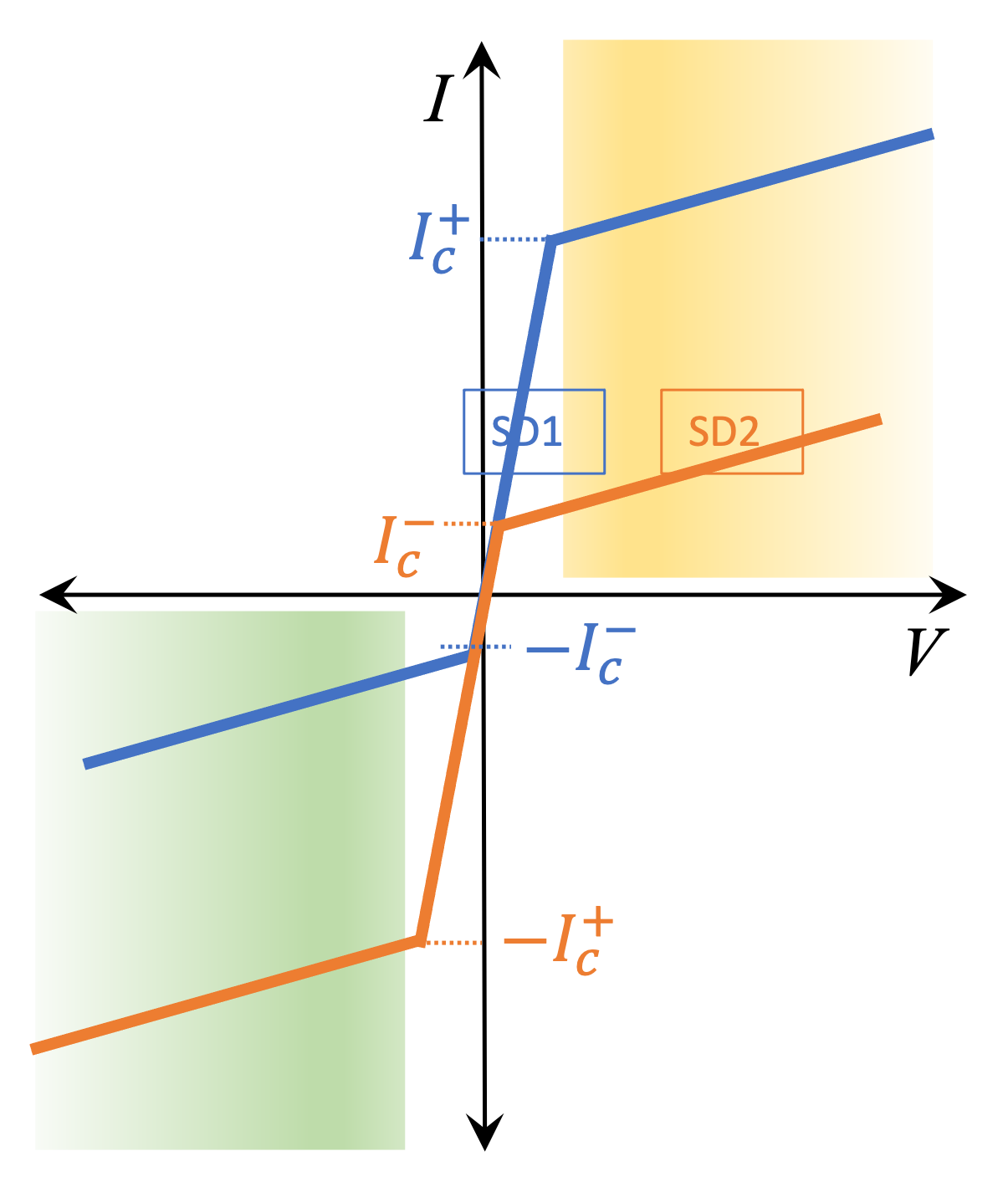}}\subfigure[]{\includegraphics[width=0.25\columnwidth]{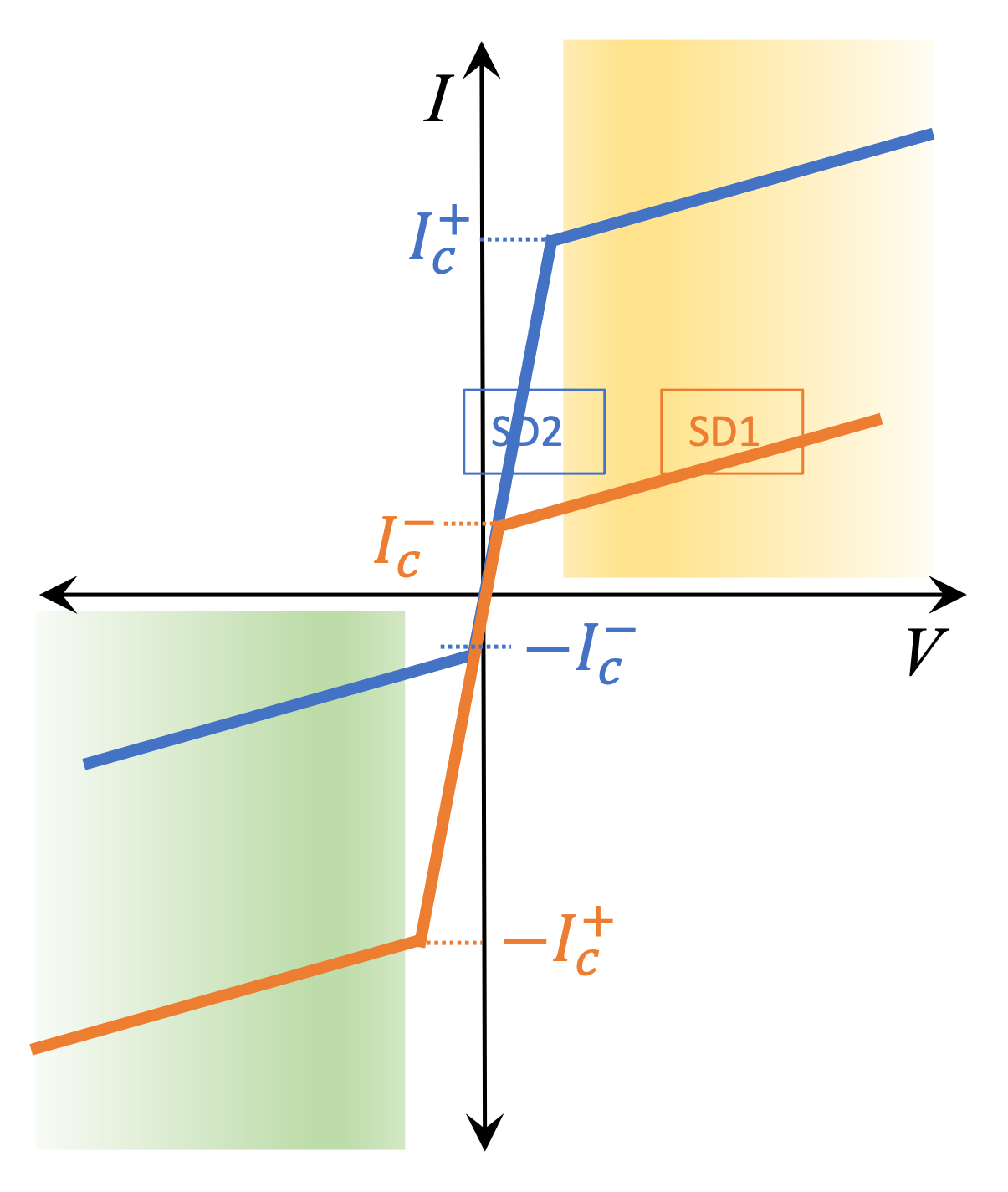}}\subfigure[]{\includegraphics[width=0.25\columnwidth]{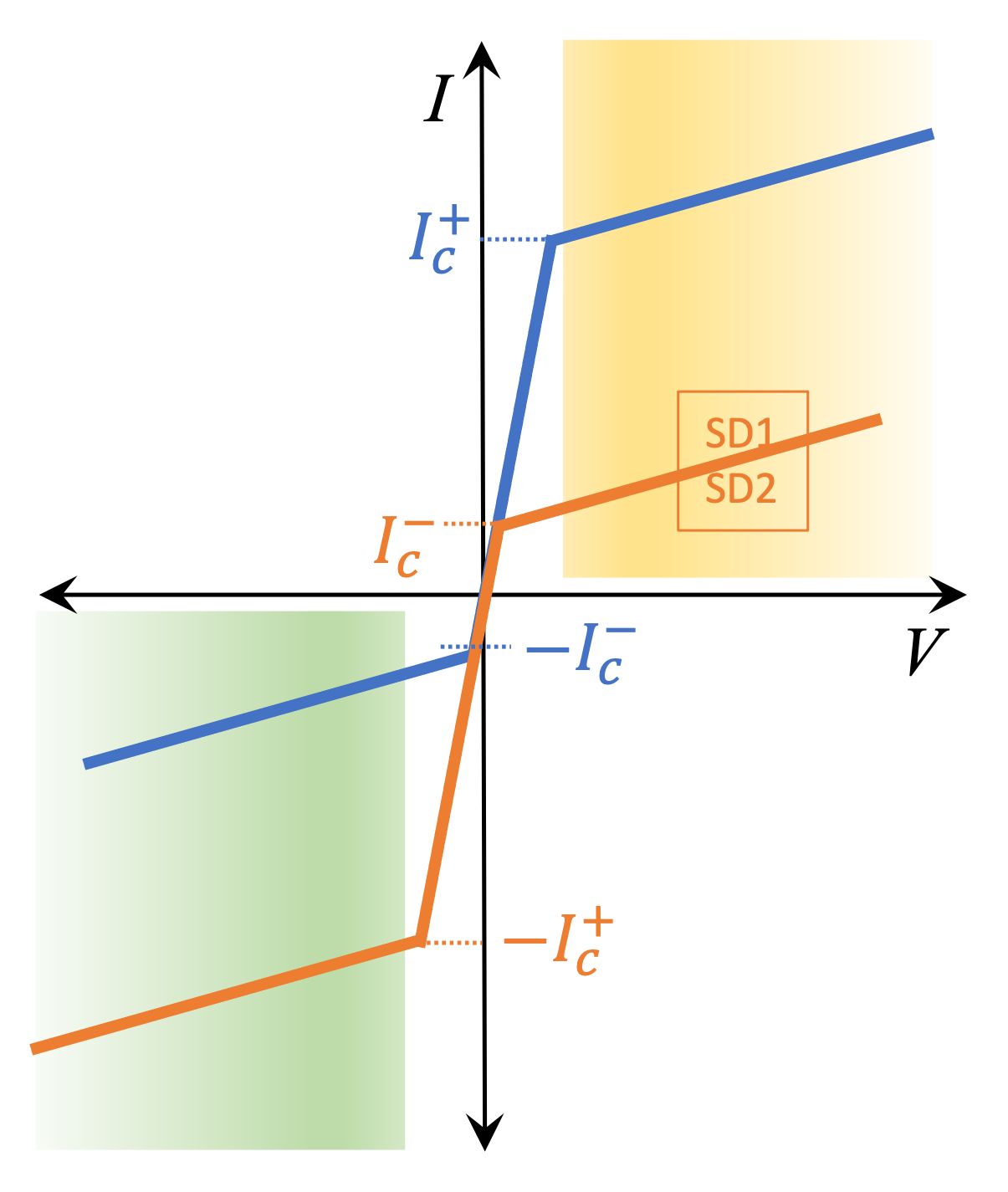}}
\par\end{centering}
\caption{Illustration of the operating regime of the SDs receiving inputs $V_{1}$
and $V_{2}$, denoted SD1 and SD2, respectively for the cases in each
of Tables \ref{tab:NOR-analysis} and \ref{tab:AND-analysis}. Subfigures
(a-d) correspond to rows (1-4) of the tables. \label{fig:SD1-SD2-states}}

\end{figure*}

For the NOR and AND gates sketched in Fig. \ref{fig:Two-bit gates}, the
circuit analysis is summarized in Tables \ref{tab:NOR-analysis} and
\ref{tab:AND-analysis}, respectively. Note that the only difference between
the two tables occurs in column two, where all the inequalities are
inverted. This reflects the simple logical fact that inverting the
inputs of NOR yields AND and vice-versa. The physical states of the
two SDs are indicated in Fig. \ref{fig:SD1-SD2-states}.

\begin{table*}
\subfigure[\label{tab:NOR-analysis}]{%
\begin{tabular}{|c|c|c|c|}
\hline 
$V_{\text{bias}}$ using Kirchoff's voltage law & Input voltages & Self-consistency condition & Physical state of SD1, SD2\tabularnewline
\hline 
\hline 
$2\left[I_{c}^{+}r+(I-I_{c}^{+})R_{N}\right]+IR$ & $V_{1}<I_{c}^{+}R/2,V_{2}<I_{c}^{+}R/2$ & $I>I_{c}^{+}$ & Normal, Normal\tabularnewline
\hline 
$Ir+I_{c}^{-}r+(I-I_{c}^{-})R_{N}+I_{c}^{-}r+IR$ & $V_{1}<I_{c}^{+}R/2,V_{2}>I_{c}^{+}R/2$ & $I_{c}^{-}<I<I_{c}^{+}$ & Superconducting, Normal\tabularnewline
\hline 
$I_{c}^{-}r+(I-I_{c}^{-})R_{N}+I_{c}^{-}r+Ir+IR$ & $V_{1}>I_{c}^{+}R/2,V_{2}<I_{c}^{+}R/2$ & $I_{c}^{-}<I<I_{c}^{+}$ & Normal, Superconducting\tabularnewline
\hline 
$2\left[I_{c}^{-}r+(I-I_{c}^{-})R_{N}\right]+IR$ & $V_{1}>I_{c}^{+}R/2,V_{2}>I_{c}^{+}R/2$ & $I_{c}^{-}<I<I_{c}^{+}$ & Normal, Normal\tabularnewline
\hline 
\end{tabular}

}

\subfigure[\label{tab:AND-analysis}]{%
\begin{tabular}{|c|c|c|c|}
\hline 
$V_{\text{bias}}$ using Kirchoff's voltage law & Input voltages & Self-consistency condition & Physical state of SD1, SD2\tabularnewline
\hline 
\hline 
$2\left[I_{c}^{+}r+(I-I_{c}^{+})R_{N}\right]+IR$ & $V_{1}>I_{c}^{+}R/2,V_{2}>I_{c}^{+}R/2$ & $I>I_{c}^{+}$ & Normal, Normal\tabularnewline
\hline 
$Ir+I_{c}^{-}r+(I-I_{c}^{-})R_{N}+I_{c}^{-}r+IR$ & $V_{1}>I_{c}^{+}R/2,V_{2}<I_{c}^{+}R/2$ & $I_{c}^{-}<I<I_{c}^{+}$ & Superconducting, Normal\tabularnewline
\hline 
$I_{c}^{-}r+(I-I_{c}^{-})R_{N}+I_{c}^{-}r+Ir+IR$ & $V_{1}<I_{c}^{+}R/2,V_{2}>I_{c}^{+}R/2$ & $I_{c}^{-}<I<I_{c}^{+}$ & Normal, Superconducting\tabularnewline
\hline 
$2\left[I_{c}^{-}r+(I-I_{c}^{-})R_{N}\right]+IR$ & $V_{1}<I_{c}^{+}R/2,V_{2}<I_{c}^{+}R/2$ & $I_{c}^{-}<I<I_{c}^{+}$ & Normal, Normal\tabularnewline
\hline 
\end{tabular}

}

\caption{Analysis of the (a) NOR and (b) AND gate circuits. SD1 and SD2 refer
to the SDs receiving inputs $V_{1}$ and $V_{2}$, respectively. For
concreteness, we restrict to a regime where the voltage drop across
each SD is positive, i.e., each SD is in the $V>0$ region per Fig.
\ref{fig:I-V}. The precise regimes of each SD for various inputs
are depicted below in Fig. \ref{fig:SD1-SD2-states}.}
\end{table*}

\bibliography{library}

\begin{thebibliography}{10}
\urlstyle{rm}
\expandafter\ifx\csname url\endcsname\relax
  \def\url#1{\texttt{#1}}\fi
\expandafter\ifx\csname urlprefix\endcsname\relax\def\urlprefix{URL }\fi
\expandafter\ifx\csname doiprefix\endcsname\relax\def\doiprefix{DOI: }\fi
\providecommand{\bibinfo}[2]{#2}
\providecommand{\eprint}[2][]{\url{#2}}

\bibitem{Reilly2019}
\bibinfo{author}{{Reilly}, D.~J.}
\newblock \bibinfo{journal}{\bibinfo{title}{{Challenges in Scaling-up the
  Control Interface of a Quantum Computer}}}.
\newblock {\emph{\JournalTitle{arXiv e-prints}}}
  \bibinfo{pages}{arXiv:1912.05114}, \doiprefix\url{10.48550/arXiv.1912.05114}
  (\bibinfo{year}{2019}).
\newblock \eprint{1912.05114}.

\bibitem{Ando:2020td}
\bibinfo{author}{Ando, F.} \emph{et~al.}
\newblock \bibinfo{journal}{\bibinfo{title}{Observation of superconducting
  diode effect}}.
\newblock {\emph{\JournalTitle{Nature}}} \textbf{\bibinfo{volume}{584}},
  \bibinfo{pages}{373--376}, \doiprefix\url{10.1038/s41586-020-2590-4}
  (\bibinfo{year}{2020}).

\bibitem{Chiles:2023aa}
\bibinfo{author}{Chiles, J.} \emph{et~al.}
\newblock \bibinfo{journal}{\bibinfo{title}{Nonreciprocal supercurrents in a
  field-free graphene josephson triode}}.
\newblock {\emph{\JournalTitle{Nano Letters}}} \textbf{\bibinfo{volume}{23}},
  \bibinfo{pages}{5257--5263}, \doiprefix\url{10.1021/acs.nanolett.3c01276}
  (\bibinfo{year}{2023}).

\bibitem{Jiang1994}
\bibinfo{author}{Jiang, X.}, \bibinfo{author}{Connolly, P.~J.},
  \bibinfo{author}{Hagen, S.~J.} \& \bibinfo{author}{Lobb, C.~J.}
\newblock \bibinfo{journal}{\bibinfo{title}{Asymmetric current-voltage
  characteristics in type-ii superconductors}}.
\newblock {\emph{\JournalTitle{Phys. Rev. B}}} \textbf{\bibinfo{volume}{49}},
  \bibinfo{pages}{9244--9247}, \doiprefix\url{10.1103/PhysRevB.49.9244}
  (\bibinfo{year}{1994}).

\bibitem{Lyu:2021wg}
\bibinfo{author}{Lyu, Y.-Y.} \emph{et~al.}
\newblock \bibinfo{journal}{\bibinfo{title}{Superconducting diode effect via
  conformal-mapped nanoholes}}.
\newblock {\emph{\JournalTitle{Nature Communications}}}
  \textbf{\bibinfo{volume}{12}}, \bibinfo{pages}{2703},
  \doiprefix\url{10.1038/s41467-021-23077-0} (\bibinfo{year}{2021}).

\bibitem{Du2023}
\bibinfo{author}{{Du}, W.-S.} \emph{et~al.}
\newblock \bibinfo{journal}{\bibinfo{title}{{Superconducting Diode Effect and
  Large Magnetochiral Anisotropy in T$_d$-MoTe$_2$ Thin Film}}}.
\newblock {\emph{\JournalTitle{arXiv e-prints}}}
  \bibinfo{pages}{arXiv:2303.09052}, \doiprefix\url{10.48550/arXiv.2303.09052}
  (\bibinfo{year}{2023}).
\newblock \eprint{2303.09052}.

\bibitem{Sundaresh:2023aa}
\bibinfo{author}{Sundaresh, A.}, \bibinfo{author}{V{\"a}yrynen, J.~I.},
  \bibinfo{author}{Lyanda-Geller, Y.} \& \bibinfo{author}{Rokhinson, L.~P.}
\newblock \bibinfo{journal}{\bibinfo{title}{Diamagnetic mechanism of critical
  current non-reciprocity in multilayered superconductors}}.
\newblock {\emph{\JournalTitle{Nature Communications}}}
  \textbf{\bibinfo{volume}{14}}, \bibinfo{pages}{1628},
  \doiprefix\url{10.1038/s41467-023-36786-5} (\bibinfo{year}{2023}).

\bibitem{Kealhofer2023}
\bibinfo{author}{Kealhofer, R.}, \bibinfo{author}{Jeong, H.},
  \bibinfo{author}{Rashidi, A.}, \bibinfo{author}{Balents, L.} \&
  \bibinfo{author}{Stemmer, S.}
\newblock \bibinfo{journal}{\bibinfo{title}{Anomalous superconducting diode
  effect in a polar superconductor}}.
\newblock {\emph{\JournalTitle{Phys. Rev. B}}} \textbf{\bibinfo{volume}{107}},
  \bibinfo{pages}{L100504}, \doiprefix\url{10.1103/PhysRevB.107.L100504}
  (\bibinfo{year}{2023}).

\bibitem{Hou2023}
\bibinfo{author}{Hou, Y.} \emph{et~al.}
\newblock \bibinfo{journal}{\bibinfo{title}{Ubiquitous superconducting diode
  effect in superconductor thin films}}.
\newblock {\emph{\JournalTitle{Phys. Rev. Lett.}}}
  \textbf{\bibinfo{volume}{131}}, \bibinfo{pages}{027001},
  \doiprefix\url{10.1103/PhysRevLett.131.027001} (\bibinfo{year}{2023}).

\bibitem{Chen:2024aa}
\bibinfo{author}{Chen, P.} \emph{et~al.}
\newblock \bibinfo{journal}{\bibinfo{title}{Edelstein effect induced
  superconducting diode effect in inversion symmetry breaking mote2 josephson
  junctions}}.
\newblock {\emph{\JournalTitle{Advanced Functional Materials}}}
  \textbf{\bibinfo{volume}{34}}, \bibinfo{pages}{2311229},
  \doiprefix\url{https://doi.org/10.1002/adfm.202311229}
  (\bibinfo{year}{2024}).

\bibitem{Gupta:2023aa}
\bibinfo{author}{Gupta, M.} \emph{et~al.}
\newblock \bibinfo{journal}{\bibinfo{title}{Gate-tunable superconducting diode
  effect in a three-terminal josephson device}}.
\newblock {\emph{\JournalTitle{Nature Communications}}}
  \textbf{\bibinfo{volume}{14}}, \bibinfo{pages}{3078},
  \doiprefix\url{10.1038/s41467-023-38856-0} (\bibinfo{year}{2023}).

\bibitem{Baumgartner_2022}
\bibinfo{author}{Baumgartner, C.} \emph{et~al.}
\newblock \bibinfo{journal}{\bibinfo{title}{Effect of rashba and dresselhaus
  spin{\textendash}orbit coupling on supercurrent rectification and
  magnetochiral anisotropy of ballistic josephson junctions}}.
\newblock {\emph{\JournalTitle{Journal of Physics: Condensed Matter}}}
  \textbf{\bibinfo{volume}{34}}, \bibinfo{pages}{154005},
  \doiprefix\url{10.1088/1361-648x/ac4d5e} (\bibinfo{year}{2022}).

\bibitem{Baumgartner:2022wr}
\bibinfo{author}{Baumgartner, C.} \emph{et~al.}
\newblock \bibinfo{journal}{\bibinfo{title}{Supercurrent rectification and
  magnetochiral effects in symmetric josephson junctions}}.
\newblock {\emph{\JournalTitle{Nature Nanotechnology}}}
  \textbf{\bibinfo{volume}{17}}, \bibinfo{pages}{39--44},
  \doiprefix\url{10.1038/s41565-021-01009-9} (\bibinfo{year}{2022}).

\bibitem{Banerjee2023}
\bibinfo{author}{Banerjee, A.} \emph{et~al.}
\newblock \bibinfo{journal}{\bibinfo{title}{Phase asymmetry of andreev spectra
  from cooper-pair momentum}}.
\newblock {\emph{\JournalTitle{Phys. Rev. Lett.}}}
  \textbf{\bibinfo{volume}{131}}, \bibinfo{pages}{196301},
  \doiprefix\url{10.1103/PhysRevLett.131.196301} (\bibinfo{year}{2023}).

\bibitem{Pal:2022tm}
\bibinfo{author}{Pal, B.} \emph{et~al.}
\newblock \bibinfo{journal}{\bibinfo{title}{Josephson diode effect from cooper
  pair momentum in a topological semimetal}}.
\newblock {\emph{\JournalTitle{Nature Physics}}}
  \doiprefix\url{10.1038/s41567-022-01699-5} (\bibinfo{year}{2022}).

\bibitem{Kim2024}
\bibinfo{author}{{Kim}, J.-K.} \emph{et~al.}
\newblock \bibinfo{journal}{\bibinfo{title}{{Intrinsic supercurrent
  non-reciprocity coupled to the crystal structure of a van der Waals Josephson
  barrier}}}.
\newblock {\emph{\JournalTitle{Nature Communications}}}
  \textbf{\bibinfo{volume}{15}}, \bibinfo{pages}{1120},
  \doiprefix\url{10.1038/s41467-024-45298-9} (\bibinfo{year}{2024}).
\newblock \eprint{2303.13049}.

\bibitem{Turini:2022aa}
\bibinfo{author}{Turini, B.} \emph{et~al.}
\newblock \bibinfo{journal}{\bibinfo{title}{Josephson diode effect in
  high-mobility insb nanoflags}}.
\newblock {\emph{\JournalTitle{Nano Letters}}} \textbf{\bibinfo{volume}{22}},
  \bibinfo{pages}{8502--8508}, \doiprefix\url{10.1021/acs.nanolett.2c02899}
  (\bibinfo{year}{2022}).

\bibitem{Bauriedl:2022we}
\bibinfo{author}{Bauriedl, L.} \emph{et~al.}
\newblock \bibinfo{journal}{\bibinfo{title}{Supercurrent diode effect and
  magnetochiral anisotropy in few-layer nbse2}}.
\newblock {\emph{\JournalTitle{Nature Communications}}}
  \textbf{\bibinfo{volume}{13}}, \bibinfo{pages}{4266},
  \doiprefix\url{10.1038/s41467-022-31954-5} (\bibinfo{year}{2022}).

\bibitem{Wu:2022wq}
\bibinfo{author}{Wu, H.} \emph{et~al.}
\newblock \bibinfo{journal}{\bibinfo{title}{The field-free josephson diode in a
  van der waals heterostructure}}.
\newblock {\emph{\JournalTitle{Nature}}} \textbf{\bibinfo{volume}{604}},
  \bibinfo{pages}{653--656}, \doiprefix\url{10.1038/s41586-022-04504-8}
  (\bibinfo{year}{2022}).

\bibitem{Diez-Merida2021}
\bibinfo{author}{Diez-Merida, J.} \emph{et~al.}
\newblock \bibinfo{title}{Magnetic josephson junctions and superconducting
  diodes in magic angle twisted bilayer graphene},
  \doiprefix\url{10.48550/ARXIV.2110.01067} (\bibinfo{year}{2021}).

\bibitem{Golod:2022ta}
\bibinfo{author}{Golod, T.} \& \bibinfo{author}{Krasnov, V.~M.}
\newblock \bibinfo{journal}{\bibinfo{title}{Demonstration of a superconducting
  diode-with-memory, operational at zero magnetic field with switchable
  nonreciprocity}}.
\newblock {\emph{\JournalTitle{Nature Communications}}}
  \textbf{\bibinfo{volume}{13}}, \bibinfo{pages}{3658},
  \doiprefix\url{10.1038/s41467-022-31256-w} (\bibinfo{year}{2022}).

\bibitem{Zhang2024}
\bibinfo{author}{Zhang, F.} \emph{et~al.}
\newblock \bibinfo{journal}{\bibinfo{title}{Magnetic-field-free nonreciprocal
  transport in graphene multiterminal josephson junctions}}.
\newblock {\emph{\JournalTitle{Phys. Rev. Appl.}}}
  \textbf{\bibinfo{volume}{21}}, \bibinfo{pages}{034011},
  \doiprefix\url{10.1103/PhysRevApplied.21.034011} (\bibinfo{year}{2024}).

\bibitem{Yun2023}
\bibinfo{author}{Yun, J.} \emph{et~al.}
\newblock \bibinfo{journal}{\bibinfo{title}{Magnetic proximity-induced
  superconducting diode effect and infinite magnetoresistance in a van der
  waals heterostructure}}.
\newblock {\emph{\JournalTitle{Phys. Rev. Res.}}} \textbf{\bibinfo{volume}{5}},
  \bibinfo{pages}{L022064}, \doiprefix\url{10.1103/PhysRevResearch.5.L022064}
  (\bibinfo{year}{2023}).

\bibitem{Jeon:2022aa}
\bibinfo{author}{Jeon, K.-R.} \emph{et~al.}
\newblock \bibinfo{journal}{\bibinfo{title}{Zero-field polarity-reversible
  josephson supercurrent diodes enabled by a proximity-magnetized pt barrier}}.
\newblock {\emph{\JournalTitle{Nature Materials}}}
  \textbf{\bibinfo{volume}{21}}, \bibinfo{pages}{1008--1013},
  \doiprefix\url{10.1038/s41563-022-01300-7} (\bibinfo{year}{2022}).

\bibitem{Lin:2022aa}
\bibinfo{author}{Lin, J.-X.} \emph{et~al.}
\newblock \bibinfo{journal}{\bibinfo{title}{Zero-field superconducting diode
  effect in small-twist-angle trilayer graphene}}.
\newblock {\emph{\JournalTitle{Nature Physics}}} \textbf{\bibinfo{volume}{18}},
  \bibinfo{pages}{1221--1227}, \doiprefix\url{10.1038/s41567-022-01700-1}
  (\bibinfo{year}{2022}).

\bibitem{Anwar:2023aa}
\bibinfo{author}{Anwar, M.~S.} \emph{et~al.}
\newblock \bibinfo{journal}{\bibinfo{title}{Spontaneous superconducting diode
  effect in non-magnetic nb/ru/sr2ruo4 topological junctions}}.
\newblock {\emph{\JournalTitle{Communications Physics}}}
  \textbf{\bibinfo{volume}{6}}, \bibinfo{pages}{290},
  \doiprefix\url{10.1038/s42005-023-01409-4} (\bibinfo{year}{2023}).

\bibitem{Narita:2022tb}
\bibinfo{author}{Narita, H.} \emph{et~al.}
\newblock \bibinfo{journal}{\bibinfo{title}{Field-free superconducting diode
  effect in noncentrosymmetric superconductor/ferromagnet multilayers}}.
\newblock {\emph{\JournalTitle{Nature Nanotechnology}}}
  \textbf{\bibinfo{volume}{17}}, \bibinfo{pages}{823--828},
  \doiprefix\url{10.1038/s41565-022-01159-4} (\bibinfo{year}{2022}).

\bibitem{Gutfreund:2023aa}
\bibinfo{author}{Gutfreund, A.} \emph{et~al.}
\newblock \bibinfo{journal}{\bibinfo{title}{Direct observation of a
  superconducting vortex diode}}.
\newblock {\emph{\JournalTitle{Nature Communications}}}
  \textbf{\bibinfo{volume}{14}}, \bibinfo{pages}{1630},
  \doiprefix\url{10.1038/s41467-023-37294-2} (\bibinfo{year}{2023}).

\bibitem{Zhao2023}
\bibinfo{author}{Zhao, S. Y.~F.} \emph{et~al.}
\newblock \bibinfo{journal}{\bibinfo{title}{Time-reversal symmetry breaking
  superconductivity between twisted cuprate superconductors}}.
\newblock {\emph{\JournalTitle{Science}}} \textbf{\bibinfo{volume}{382}},
  \bibinfo{pages}{1422--1427}, \doiprefix\url{10.1126/science.abl8371}
  (\bibinfo{year}{2023}).
\newblock \eprint{https://www.science.org/doi/pdf/10.1126/science.abl8371}.

\bibitem{Trahms:2023aa}
\bibinfo{author}{Trahms, M.} \emph{et~al.}
\newblock \bibinfo{journal}{\bibinfo{title}{Diode effect in josephson junctions
  with a single magnetic atom}}.
\newblock {\emph{\JournalTitle{Nature}}} \textbf{\bibinfo{volume}{615}},
  \bibinfo{pages}{628--633}, \doiprefix\url{10.1038/s41586-023-05743-z}
  (\bibinfo{year}{2023}).

\bibitem{Yasuda:2019wb}
\bibinfo{author}{Yasuda, K.} \emph{et~al.}
\newblock \bibinfo{journal}{\bibinfo{title}{Nonreciprocal charge transport at
  topological insulator/superconductor interface}}.
\newblock {\emph{\JournalTitle{Nature Communications}}}
  \textbf{\bibinfo{volume}{10}}, \bibinfo{pages}{2734},
  \doiprefix\url{10.1038/s41467-019-10658-3} (\bibinfo{year}{2019}).

\bibitem{Masuko:2022aa}
\bibinfo{author}{Masuko, M.} \emph{et~al.}
\newblock \bibinfo{journal}{\bibinfo{title}{Nonreciprocal charge transport in
  topological superconductor candidate bi2te3/pdte2 heterostructure}}.
\newblock {\emph{\JournalTitle{npj Quantum Materials}}}
  \textbf{\bibinfo{volume}{7}}, \bibinfo{pages}{104},
  \doiprefix\url{10.1038/s41535-022-00514-x} (\bibinfo{year}{2022}).

\bibitem{he2024observationsuperconductingdiodeeffect}
\bibinfo{author}{He, J.} \emph{et~al.}
\newblock \bibinfo{title}{Observation of superconducting diode effect in
  antiferromagnetic mott insulator $\alpha$-rucl$_3$} (\bibinfo{year}{2024}).
\newblock \eprint{2409.04093}.

\bibitem{Anh:2024aa}
\bibinfo{author}{Anh, L.~D.} \emph{et~al.}
\newblock \bibinfo{journal}{\bibinfo{title}{Large superconducting diode effect
  in ion-beam patterned sn-based superconductor nanowire/topological dirac
  semimetal planar heterostructures}}.
\newblock {\emph{\JournalTitle{Nature Communications}}}
  \textbf{\bibinfo{volume}{15}}, \bibinfo{pages}{8014},
  \doiprefix\url{10.1038/s41467-024-52080-4} (\bibinfo{year}{2024}).

\bibitem{Yuan2022}
\bibinfo{author}{Yuan, N. F.~Q.} \& \bibinfo{author}{Fu, L.}
\newblock \bibinfo{journal}{\bibinfo{title}{Supercurrent diode effect and
  finite-momentum superconductors}}.
\newblock {\emph{\JournalTitle{Proceedings of the National Academy of
  Sciences}}} \textbf{\bibinfo{volume}{119}}, \bibinfo{pages}{e2119548119},
  \doiprefix\url{10.1073/pnas.2119548119} (\bibinfo{year}{2022}).
\newblock \eprint{https://www.pnas.org/doi/pdf/10.1073/pnas.2119548119}.

\bibitem{Daido2022}
\bibinfo{author}{Daido, A.}, \bibinfo{author}{Ikeda, Y.} \&
  \bibinfo{author}{Yanase, Y.}
\newblock \bibinfo{journal}{\bibinfo{title}{Intrinsic superconducting diode
  effect}}.
\newblock {\emph{\JournalTitle{Phys. Rev. Lett.}}}
  \textbf{\bibinfo{volume}{128}}, \bibinfo{pages}{037001},
  \doiprefix\url{10.1103/PhysRevLett.128.037001} (\bibinfo{year}{2022}).

\bibitem{Daido2022a}
\bibinfo{author}{Daido, A.} \& \bibinfo{author}{Yanase, Y.}
\newblock \bibinfo{journal}{\bibinfo{title}{Superconducting diode effect and
  nonreciprocal transition lines}}.
\newblock {\emph{\JournalTitle{Phys. Rev. B}}} \textbf{\bibinfo{volume}{106}},
  \bibinfo{pages}{205206}, \doiprefix\url{10.1103/PhysRevB.106.205206}
  (\bibinfo{year}{2022}).

\bibitem{Zhang2022}
\bibinfo{author}{Zhang, Y.}, \bibinfo{author}{Gu, Y.}, \bibinfo{author}{Li,
  P.}, \bibinfo{author}{Hu, J.} \& \bibinfo{author}{Jiang, K.}
\newblock \bibinfo{journal}{\bibinfo{title}{General theory of josephson
  diodes}}.
\newblock {\emph{\JournalTitle{Phys. Rev. X}}} \textbf{\bibinfo{volume}{12}},
  \bibinfo{pages}{041013}, \doiprefix\url{10.1103/PhysRevX.12.041013}
  (\bibinfo{year}{2022}).

\bibitem{Davydova2022}
\bibinfo{author}{Davydova, M.}, \bibinfo{author}{Prembabu, S.} \&
  \bibinfo{author}{Fu, L.}
\newblock \bibinfo{journal}{\bibinfo{title}{Universal josephson diode effect}}.
\newblock {\emph{\JournalTitle{Science Advances}}}
  \textbf{\bibinfo{volume}{8}}, \bibinfo{pages}{eabo0309},
  \doiprefix\url{10.1126/sciadv.abo0309} (\bibinfo{year}{2022}).
\newblock \eprint{https://www.science.org/doi/pdf/10.1126/sciadv.abo0309}.

\bibitem{Chen2024}
\bibinfo{author}{Chen, K.}, \bibinfo{author}{Karki, B.} \&
  \bibinfo{author}{Hosur, P.}
\newblock \bibinfo{journal}{\bibinfo{title}{Intrinsic superconducting diode
  effects in tilted weyl and dirac semimetals}}.
\newblock {\emph{\JournalTitle{Phys. Rev. B}}} \textbf{\bibinfo{volume}{109}},
  \bibinfo{pages}{064511}, \doiprefix\url{10.1103/PhysRevB.109.064511}
  (\bibinfo{year}{2024}).

\bibitem{wang2022symmetry}
\bibinfo{author}{Wang, D.}, \bibinfo{author}{Wang, Q.-H.} \&
  \bibinfo{author}{Wu, C.}
\newblock \bibinfo{title}{Symmetry constraints on direct-current josephson
  diodes} (\bibinfo{year}{2022}).
\newblock \eprint{2209.12646}.

\bibitem{He_2022}
\bibinfo{author}{He, J.~J.}, \bibinfo{author}{Tanaka, Y.} \&
  \bibinfo{author}{Nagaosa, N.}
\newblock \bibinfo{journal}{\bibinfo{title}{A phenomenological theory of
  superconductor diodes}}.
\newblock {\emph{\JournalTitle{New Journal of Physics}}}
  \textbf{\bibinfo{volume}{24}}, \bibinfo{pages}{053014},
  \doiprefix\url{10.1088/1367-2630/ac6766} (\bibinfo{year}{2022}).

\bibitem{Zhai2022}
\bibinfo{author}{Zhai, B.}, \bibinfo{author}{Li, B.}, \bibinfo{author}{Wen,
  Y.}, \bibinfo{author}{Wu, F.} \& \bibinfo{author}{He, J.}
\newblock \bibinfo{journal}{\bibinfo{title}{Prediction of ferroelectric
  superconductors with reversible superconducting diode effect}}.
\newblock {\emph{\JournalTitle{Phys. Rev. B}}} \textbf{\bibinfo{volume}{106}},
  \bibinfo{pages}{L140505}, \doiprefix\url{10.1103/PhysRevB.106.L140505}
  (\bibinfo{year}{2022}).

\bibitem{MIsaki2021}
\bibinfo{author}{Misaki, K.} \& \bibinfo{author}{Nagaosa, N.}
\newblock \bibinfo{journal}{\bibinfo{title}{Theory of the nonreciprocal
  josephson effect}}.
\newblock {\emph{\JournalTitle{Phys. Rev. B}}} \textbf{\bibinfo{volume}{103}},
  \bibinfo{pages}{245302}, \doiprefix\url{10.1103/PhysRevB.103.245302}
  (\bibinfo{year}{2021}).

\bibitem{Ilic2022}
\bibinfo{author}{Ili\ifmmode~\acute{c}\else \'{c}\fi{}, S.} \&
  \bibinfo{author}{Bergeret, F.~S.}
\newblock \bibinfo{journal}{\bibinfo{title}{Theory of the supercurrent diode
  effect in rashba superconductors with arbitrary disorder}}.
\newblock {\emph{\JournalTitle{Phys. Rev. Lett.}}}
  \textbf{\bibinfo{volume}{128}}, \bibinfo{pages}{177001},
  \doiprefix\url{10.1103/PhysRevLett.128.177001} (\bibinfo{year}{2022}).

\bibitem{Scammell_2022}
\bibinfo{author}{Scammell, H.~D.}, \bibinfo{author}{Li, J. I.~A.} \&
  \bibinfo{author}{Scheurer, M.~S.}
\newblock \bibinfo{journal}{\bibinfo{title}{Theory of zero-field
  superconducting diode effect in twisted trilayer graphene}}.
\newblock {\emph{\JournalTitle{2D Materials}}} \textbf{\bibinfo{volume}{9}},
  \bibinfo{pages}{025027}, \doiprefix\url{10.1088/2053-1583/ac5b16}
  (\bibinfo{year}{2022}).

\bibitem{Zinkl2022}
\bibinfo{author}{Zinkl, B.}, \bibinfo{author}{Hamamoto, K.} \&
  \bibinfo{author}{Sigrist, M.}
\newblock \bibinfo{journal}{\bibinfo{title}{Symmetry conditions for the
  superconducting diode effect in chiral superconductors}}.
\newblock {\emph{\JournalTitle{Phys. Rev. Res.}}} \textbf{\bibinfo{volume}{4}},
  \bibinfo{pages}{033167}, \doiprefix\url{10.1103/PhysRevResearch.4.033167}
  (\bibinfo{year}{2022}).

\bibitem{He:2023aa}
\bibinfo{author}{He, J.~J.}, \bibinfo{author}{Tanaka, Y.} \&
  \bibinfo{author}{Nagaosa, N.}
\newblock \bibinfo{journal}{\bibinfo{title}{The supercurrent diode effect and
  nonreciprocal paraconductivity due to the chiral structure of nanotubes}}.
\newblock {\emph{\JournalTitle{Nature Communications}}}
  \textbf{\bibinfo{volume}{14}}, \bibinfo{pages}{3330},
  \doiprefix\url{10.1038/s41467-023-39083-3} (\bibinfo{year}{2023}).

\bibitem{Jiang2022}
\bibinfo{author}{Jiang, J.} \emph{et~al.}
\newblock \bibinfo{journal}{\bibinfo{title}{Field-free superconducting diode in
  a magnetically nanostructured superconductor}}.
\newblock {\emph{\JournalTitle{Phys. Rev. Appl.}}}
  \textbf{\bibinfo{volume}{18}}, \bibinfo{pages}{034064},
  \doiprefix\url{10.1103/PhysRevApplied.18.034064} (\bibinfo{year}{2022}).

\bibitem{Kokkeler2022}
\bibinfo{author}{Kokkeler, T.~H.}, \bibinfo{author}{Golubov, A.~A.} \&
  \bibinfo{author}{Bergeret, F.~S.}
\newblock \bibinfo{journal}{\bibinfo{title}{Field-free anomalous junction and
  superconducting diode effect in spin-split superconductor/topological
  insulator junctions}}.
\newblock {\emph{\JournalTitle{Phys. Rev. B}}} \textbf{\bibinfo{volume}{106}},
  \bibinfo{pages}{214504}, \doiprefix\url{10.1103/PhysRevB.106.214504}
  (\bibinfo{year}{2022}).

\bibitem{Debnath2024}
\bibinfo{author}{Debnath, D.} \& \bibinfo{author}{Dutta, P.}
\newblock \bibinfo{journal}{\bibinfo{title}{Gate-tunable josephson diode effect
  in rashba spin-orbit coupled quantum dot junctions}}.
\newblock {\emph{\JournalTitle{Phys. Rev. B}}} \textbf{\bibinfo{volume}{109}},
  \bibinfo{pages}{174511}, \doiprefix\url{10.1103/PhysRevB.109.174511}
  (\bibinfo{year}{2024}).

\bibitem{Chazono2023}
\bibinfo{author}{Chazono, M.}, \bibinfo{author}{Kanasugi, S.},
  \bibinfo{author}{Kitamura, T.} \& \bibinfo{author}{Yanase, Y.}
\newblock \bibinfo{journal}{\bibinfo{title}{Piezoelectric effect and diode
  effect in anapole and monopole superconductors}}.
\newblock {\emph{\JournalTitle{Phys. Rev. B}}} \textbf{\bibinfo{volume}{107}},
  \bibinfo{pages}{214512}, \doiprefix\url{10.1103/PhysRevB.107.214512}
  (\bibinfo{year}{2023}).

\bibitem{Vodolazov2005}
\bibinfo{author}{Vodolazov, D.~Y.} \& \bibinfo{author}{Peeters, F.~M.}
\newblock \bibinfo{journal}{\bibinfo{title}{Superconducting rectifier based on
  the asymmetric surface barrier effect}}.
\newblock {\emph{\JournalTitle{Phys. Rev. B}}} \textbf{\bibinfo{volume}{72}},
  \bibinfo{pages}{172508}, \doiprefix\url{10.1103/PhysRevB.72.172508}
  (\bibinfo{year}{2005}).

\bibitem{dePicoli2023}
\bibinfo{author}{de~Picoli, T.}, \bibinfo{author}{Blood, Z.},
  \bibinfo{author}{Lyanda-Geller, Y.} \& \bibinfo{author}{V\"ayrynen, J.~I.}
\newblock \bibinfo{journal}{\bibinfo{title}{Superconducting diode effect in
  quasi-one-dimensional systems}}.
\newblock {\emph{\JournalTitle{Phys. Rev. B}}} \textbf{\bibinfo{volume}{107}},
  \bibinfo{pages}{224518}, \doiprefix\url{10.1103/PhysRevB.107.224518}
  (\bibinfo{year}{2023}).

\bibitem{Kochan2023}
\bibinfo{author}{{Kochan}, D.}, \bibinfo{author}{{Costa}, A.},
  \bibinfo{author}{{Zhumagulov}, I.} \& \bibinfo{author}{{{\v{Z}}uti{\'c}}, I.}
\newblock \bibinfo{journal}{\bibinfo{title}{{Phenomenological Theory of the
  Supercurrent Diode Effect: The Lifshitz Invariant}}}.
\newblock {\emph{\JournalTitle{arXiv e-prints}}}
  \bibinfo{pages}{arXiv:2303.11975}, \doiprefix\url{10.48550/arXiv.2303.11975}
  (\bibinfo{year}{2023}).
\newblock \eprint{2303.11975}.

\bibitem{Ikeda2022}
\bibinfo{author}{{Ikeda}, Y.}, \bibinfo{author}{{Daido}, A.} \&
  \bibinfo{author}{{Yanase}, Y.}
\newblock \bibinfo{journal}{\bibinfo{title}{{Intrinsic superconducting diode
  effect in disordered systems}}}.
\newblock {\emph{\JournalTitle{arXiv e-prints}}}
  \bibinfo{pages}{arXiv:2212.09211}, \doiprefix\url{10.48550/arXiv.2212.09211}
  (\bibinfo{year}{2022}).
\newblock \eprint{2212.09211}.

\bibitem{Tanaka2022}
\bibinfo{author}{Tanaka, Y.}, \bibinfo{author}{Lu, B.} \&
  \bibinfo{author}{Nagaosa, N.}
\newblock \bibinfo{journal}{\bibinfo{title}{Theory of giant diode effect in
  $d$-wave superconductor junctions on the surface of a topological
  insulator}}.
\newblock {\emph{\JournalTitle{Phys. Rev. B}}} \textbf{\bibinfo{volume}{106}},
  \bibinfo{pages}{214524}, \doiprefix\url{10.1103/PhysRevB.106.214524}
  (\bibinfo{year}{2022}).

\bibitem{Wang2022}
\bibinfo{author}{{Wang}, D.}, \bibinfo{author}{{Wang}, Q.-H.} \&
  \bibinfo{author}{{Wu}, C.}
\newblock \bibinfo{journal}{\bibinfo{title}{{Symmetry Constraints on
  Direct-Current Josephson Diodes}}}.
\newblock {\emph{\JournalTitle{arXiv e-prints}}}
  \bibinfo{pages}{arXiv:2209.12646}, \doiprefix\url{10.48550/arXiv.2209.12646}
  (\bibinfo{year}{2022}).
\newblock \eprint{2209.12646}.

\bibitem{Haenel2022}
\bibinfo{author}{{Haenel}, R.} \& \bibinfo{author}{{Can}, O.}
\newblock \bibinfo{journal}{\bibinfo{title}{{Superconducting diode from flux
  biased Josephson junction arrays}}}.
\newblock {\emph{\JournalTitle{arXiv e-prints}}}
  \bibinfo{pages}{arXiv:2212.02657}, \doiprefix\url{10.48550/arXiv.2212.02657}
  (\bibinfo{year}{2022}).
\newblock \eprint{2212.02657}.

\bibitem{Legg2023}
\bibinfo{author}{Legg, H.~F.}, \bibinfo{author}{Laubscher, K.},
  \bibinfo{author}{Loss, D.} \& \bibinfo{author}{Klinovaja, J.}
\newblock \bibinfo{journal}{\bibinfo{title}{Parity-protected superconducting
  diode effect in topological josephson junctions}}.
\newblock {\emph{\JournalTitle{Phys. Rev. B}}} \textbf{\bibinfo{volume}{108}},
  \bibinfo{pages}{214520}, \doiprefix\url{10.1103/PhysRevB.108.214520}
  (\bibinfo{year}{2023}).

\bibitem{Cuozzo2024}
\bibinfo{author}{Cuozzo, J.~J.}, \bibinfo{author}{Pan, W.},
  \bibinfo{author}{Shabani, J.} \& \bibinfo{author}{Rossi, E.}
\newblock \bibinfo{journal}{\bibinfo{title}{Microwave-tunable diode effect in
  asymmetric squids with topological josephson junctions}}.
\newblock {\emph{\JournalTitle{Phys. Rev. Res.}}} \textbf{\bibinfo{volume}{6}},
  \bibinfo{pages}{023011}, \doiprefix\url{10.1103/PhysRevResearch.6.023011}
  (\bibinfo{year}{2024}).

\bibitem{Suoto2022}
\bibinfo{author}{Souto, R.~S.}, \bibinfo{author}{Leijnse, M.} \&
  \bibinfo{author}{Schrade, C.}
\newblock \bibinfo{journal}{\bibinfo{title}{Josephson diode effect in
  supercurrent interferometers}}.
\newblock {\emph{\JournalTitle{Phys. Rev. Lett.}}}
  \textbf{\bibinfo{volume}{129}}, \bibinfo{pages}{267702},
  \doiprefix\url{10.1103/PhysRevLett.129.267702} (\bibinfo{year}{2022}).

\bibitem{Suoto2024}
\bibinfo{author}{Seoane~Souto, R.} \emph{et~al.}
\newblock \bibinfo{journal}{\bibinfo{title}{Tuning the josephson diode response
  with an ac current}}.
\newblock {\emph{\JournalTitle{Phys. Rev. Res.}}} \textbf{\bibinfo{volume}{6}},
  \bibinfo{pages}{L022002}, \doiprefix\url{10.1103/PhysRevResearch.6.L022002}
  (\bibinfo{year}{2024}).

\bibitem{Cheng2023}
\bibinfo{author}{Cheng, Q.} \& \bibinfo{author}{Sun, Q.-F.}
\newblock \bibinfo{journal}{\bibinfo{title}{Josephson diode based on
  conventional superconductors and a chiral quantum dot}}.
\newblock {\emph{\JournalTitle{Phys. Rev. B}}} \textbf{\bibinfo{volume}{107}},
  \bibinfo{pages}{184511}, \doiprefix\url{10.1103/PhysRevB.107.184511}
  (\bibinfo{year}{2023}).

\bibitem{Steiner2023}
\bibinfo{author}{Steiner, J.~F.}, \bibinfo{author}{Melischek, L.},
  \bibinfo{author}{Trahms, M.}, \bibinfo{author}{Franke, K.~J.} \&
  \bibinfo{author}{von Oppen, F.}
\newblock \bibinfo{journal}{\bibinfo{title}{Diode effects in current-biased
  josephson junctions}}.
\newblock {\emph{\JournalTitle{Phys. Rev. Lett.}}}
  \textbf{\bibinfo{volume}{130}}, \bibinfo{pages}{177002},
  \doiprefix\url{10.1103/PhysRevLett.130.177002} (\bibinfo{year}{2023}).

\bibitem{Costa2023}
\bibinfo{author}{Costa, A.}, \bibinfo{author}{Fabian, J.} \&
  \bibinfo{author}{Kochan, D.}
\newblock \bibinfo{journal}{\bibinfo{title}{Microscopic study of the josephson
  supercurrent diode effect in josephson junctions based on two-dimensional
  electron gas}}.
\newblock {\emph{\JournalTitle{Phys. Rev. B}}} \textbf{\bibinfo{volume}{108}},
  \bibinfo{pages}{054522}, \doiprefix\url{10.1103/PhysRevB.108.054522}
  (\bibinfo{year}{2023}).

\bibitem{Wei2022}
\bibinfo{author}{Wei, Y.-J.}, \bibinfo{author}{Liu, H.-L.},
  \bibinfo{author}{Wang, J.} \& \bibinfo{author}{Liu, J.-F.}
\newblock \bibinfo{journal}{\bibinfo{title}{Supercurrent rectification effect
  in graphene-based josephson junctions}}.
\newblock {\emph{\JournalTitle{Phys. Rev. B}}} \textbf{\bibinfo{volume}{106}},
  \bibinfo{pages}{165419}, \doiprefix\url{10.1103/PhysRevB.106.165419}
  (\bibinfo{year}{2022}).

\bibitem{Legg2022}
\bibinfo{author}{Legg, H.~F.}, \bibinfo{author}{Loss, D.} \&
  \bibinfo{author}{Klinovaja, J.}
\newblock \bibinfo{journal}{\bibinfo{title}{Superconducting diode effect due to
  magnetochiral anisotropy in topological insulators and rashba nanowires}}.
\newblock {\emph{\JournalTitle{Phys. Rev. B}}} \textbf{\bibinfo{volume}{106}},
  \bibinfo{pages}{104501}, \doiprefix\url{10.1103/PhysRevB.106.104501}
  (\bibinfo{year}{2022}).

\bibitem{Karabassov2022}
\bibinfo{author}{Karabassov, T.}, \bibinfo{author}{Bobkova, I.~V.},
  \bibinfo{author}{Golubov, A.~A.} \& \bibinfo{author}{Vasenko, A.~S.}
\newblock \bibinfo{journal}{\bibinfo{title}{Hybrid helical state and
  superconducting diode effect in superconductor/ferromagnet/topological
  insulator heterostructures}}.
\newblock {\emph{\JournalTitle{Phys. Rev. B}}} \textbf{\bibinfo{volume}{106}},
  \bibinfo{pages}{224509}, \doiprefix\url{10.1103/PhysRevB.106.224509}
  (\bibinfo{year}{2022}).

\bibitem{Hu2023}
\bibinfo{author}{Hu, J.-X.}, \bibinfo{author}{Sun, Z.-T.},
  \bibinfo{author}{Xie, Y.-M.} \& \bibinfo{author}{Law, K.~T.}
\newblock \bibinfo{journal}{\bibinfo{title}{Josephson diode effect induced by
  valley polarization in twisted bilayer graphene}}.
\newblock {\emph{\JournalTitle{Phys. Rev. Lett.}}}
  \textbf{\bibinfo{volume}{130}}, \bibinfo{pages}{266003},
  \doiprefix\url{10.1103/PhysRevLett.130.266003} (\bibinfo{year}{2023}).

\bibitem{Wu2023}
\bibinfo{author}{Wu, Y.-M.}, \bibinfo{author}{Wu, Z.} \& \bibinfo{author}{Yao,
  H.}
\newblock \bibinfo{journal}{\bibinfo{title}{Pair-density-wave and chiral
  superconductivity in twisted bilayer transition metal dichalcogenides}}.
\newblock {\emph{\JournalTitle{Phys. Rev. Lett.}}}
  \textbf{\bibinfo{volume}{130}}, \bibinfo{pages}{126001},
  \doiprefix\url{10.1103/PhysRevLett.130.126001} (\bibinfo{year}{2023}).

\bibitem{Nunchot2024}
\bibinfo{author}{Nunchot, N.} \& \bibinfo{author}{Yanase, Y.}
\newblock \bibinfo{journal}{\bibinfo{title}{Chiral superconducting diode effect
  by dzyaloshinsky-moriya interaction}}.
\newblock {\emph{\JournalTitle{Phys. Rev. B}}} \textbf{\bibinfo{volume}{109}},
  \bibinfo{pages}{054508}, \doiprefix\url{10.1103/PhysRevB.109.054508}
  (\bibinfo{year}{2024}).

\bibitem{Daido2023}
\bibinfo{author}{{Daido}, A.} \& \bibinfo{author}{{Yanase}, Y.}
\newblock \bibinfo{journal}{\bibinfo{title}{{Unidirectional Superconductivity
  and Diode Effect Induced by Dissipation}}}.
\newblock {\emph{\JournalTitle{arXiv e-prints}}}
  \bibinfo{pages}{arXiv:2310.02539}, \doiprefix\url{10.48550/arXiv.2310.02539}
  (\bibinfo{year}{2023}).
\newblock \eprint{2310.02539}.

\bibitem{Cayao2024}
\bibinfo{author}{Cayao, J.}, \bibinfo{author}{Nagaosa, N.} \&
  \bibinfo{author}{Tanaka, Y.}
\newblock \bibinfo{journal}{\bibinfo{title}{Enhancing the josephson diode
  effect with majorana bound states}}.
\newblock {\emph{\JournalTitle{Phys. Rev. B}}} \textbf{\bibinfo{volume}{109}},
  \bibinfo{pages}{L081405}, \doiprefix\url{10.1103/PhysRevB.109.L081405}
  (\bibinfo{year}{2024}).

\bibitem{Banerjee2024}
\bibinfo{author}{Banerjee, S.} \& \bibinfo{author}{Scheurer, M.~S.}
\newblock \bibinfo{journal}{\bibinfo{title}{Enhanced superconducting diode
  effect due to coexisting phases}}.
\newblock {\emph{\JournalTitle{Phys. Rev. Lett.}}}
  \textbf{\bibinfo{volume}{132}}, \bibinfo{pages}{046003},
  \doiprefix\url{10.1103/PhysRevLett.132.046003} (\bibinfo{year}{2024}).

\bibitem{Hosur2023}
\bibinfo{author}{Hosur, P.} \& \bibinfo{author}{Palacios, D.}
\newblock \bibinfo{journal}{\bibinfo{title}{Proximity-induced equilibrium
  supercurrent and perfect superconducting diode effect due to band
  asymmetry}}.
\newblock {\emph{\JournalTitle{Phys. Rev. B}}} \textbf{\bibinfo{volume}{108}},
  \bibinfo{pages}{094513}, \doiprefix\url{10.1103/PhysRevB.108.094513}
  (\bibinfo{year}{2023}).

\bibitem{chakraborty2024perfectsuperconductingdiodeeffect}
\bibinfo{author}{Chakraborty, D.} \& \bibinfo{author}{Black-Schaffer, A.~M.}
\newblock \bibinfo{title}{Perfect superconducting diode effect in altermagnets}
  (\bibinfo{year}{2024}).
\newblock \eprint{2408.07747}.

\bibitem{Soori2024}
\bibinfo{author}{{Soori}, A.}
\newblock \bibinfo{journal}{\bibinfo{title}{{Josephson diode effect in
  one-dimensional quantum wires connected to superconductors with mixed
  singlet-triplet pairing}}}.
\newblock {\emph{\JournalTitle{arXiv e-prints}}}
  \bibinfo{pages}{arXiv:2409.02794}, \doiprefix\url{10.48550/arXiv.2409.02794}
  (\bibinfo{year}{2024}).
\newblock \eprint{2409.02794}.

\bibitem{Soori_2024}
\bibinfo{author}{Soori, A.}
\newblock \bibinfo{journal}{\bibinfo{title}{Josephson diode effect in junctions
  of superconductors with band asymmetric metals}}.
\newblock {\emph{\JournalTitle{Journal of Physics: Condensed Matter}}}
  \textbf{\bibinfo{volume}{36}}, \bibinfo{pages}{335303},
  \doiprefix\url{10.1088/1361-648X/ad4aad} (\bibinfo{year}{2024}).

\bibitem{Watanabe:2019aa}
\bibinfo{author}{Watanabe, H.}
\newblock \bibinfo{journal}{\bibinfo{title}{A proof of the bloch theorem for
  lattice models}}.
\newblock {\emph{\JournalTitle{Journal of Statistical Physics}}}
  \textbf{\bibinfo{volume}{177}}, \bibinfo{pages}{717--726},
  \doiprefix\url{10.1007/s10955-019-02386-1} (\bibinfo{year}{2019}).

\bibitem{yuan2023surfacesupercurrentdiodeeffect}
\bibinfo{author}{Yuan, N. F.~Q.}
\newblock \bibinfo{title}{Surface supercurrent diode effect}
  (\bibinfo{year}{2023}).
\newblock \eprint{2305.04219}.

\bibitem{Ingla-Aynes2024}
\bibinfo{author}{{Ingla-Ayn{\'e}s}, J.} \emph{et~al.}
\newblock \bibinfo{journal}{\bibinfo{title}{{Highly Efficient Superconducting
  Diodes and Rectifiers for Quantum Circuitry}}}.
\newblock {\emph{\JournalTitle{arXiv e-prints}}}
  \bibinfo{pages}{arXiv:2406.12012}, \doiprefix\url{10.48550/arXiv.2406.12012}
  (\bibinfo{year}{2024}).
\newblock \eprint{2406.12012}.

\bibitem{shin2024electriccontrolpolarityspinorbit}
\bibinfo{author}{Shin, J.} \emph{et~al.}
\newblock \bibinfo{title}{Electric control of polarity in spin-orbit josephson
  diode} (\bibinfo{year}{2024}).
\newblock \eprint{2409.17820}.

\bibitem{Anderson1965}
\bibinfo{author}{Anderson, P.~W.} \& \bibinfo{author}{Blount, E.~I.}
\newblock \bibinfo{journal}{\bibinfo{title}{Symmetry considerations on
  martensitic transformations: "ferroelectric" metals?}}
\newblock {\emph{\JournalTitle{Phys. Rev. Lett.}}}
  \textbf{\bibinfo{volume}{14}}, \bibinfo{pages}{217--219},
  \doiprefix\url{10.1103/PhysRevLett.14.217} (\bibinfo{year}{1965}).

\bibitem{Fei:2018aa}
\bibinfo{author}{Fei, Z.} \emph{et~al.}
\newblock \bibinfo{journal}{\bibinfo{title}{Ferroelectric switching of a
  two-dimensional metal}}.
\newblock {\emph{\JournalTitle{Nature}}} \textbf{\bibinfo{volume}{560}},
  \bibinfo{pages}{336--339}, \doiprefix\url{10.1038/s41586-018-0336-3}
  (\bibinfo{year}{2018}).

\bibitem{Barrera:2021tb}
\bibinfo{author}{de~la Barrera, S.~C.} \emph{et~al.}
\newblock \bibinfo{journal}{\bibinfo{title}{Direct measurement of ferroelectric
  polarization in a tunable semimetal}}.
\newblock {\emph{\JournalTitle{Nature Communications}}}
  \textbf{\bibinfo{volume}{12}}, \bibinfo{pages}{5298},
  \doiprefix\url{10.1038/s41467-021-25587-3} (\bibinfo{year}{2021}).

\bibitem{Zheng:2020ui}
\bibinfo{author}{Zheng, Z.} \emph{et~al.}
\newblock \bibinfo{journal}{\bibinfo{title}{Unconventional ferroelectricity in
  moir{\'e}heterostructures}}.
\newblock {\emph{\JournalTitle{Nature}}} \textbf{\bibinfo{volume}{588}},
  \bibinfo{pages}{71--76}, \doiprefix\url{10.1038/s41586-020-2970-9}
  (\bibinfo{year}{2020}).

\bibitem{Jindal:2023uh}
\bibinfo{author}{Jindal, A.} \emph{et~al.}
\newblock \bibinfo{journal}{\bibinfo{title}{Coupled ferroelectricity and
  superconductivity in bilayer td-mote2}}.
\newblock {\emph{\JournalTitle{Nature}}} \textbf{\bibinfo{volume}{613}},
  \bibinfo{pages}{48--52}, \doiprefix\url{10.1038/s41586-022-05521-3}
  (\bibinfo{year}{2023}).

\bibitem{ZhangFCGT}
\bibinfo{author}{Zhang, H.} \emph{et~al.}
\newblock \bibinfo{journal}{\bibinfo{title}{A room temperature polar magnetic
  metal}}.
\newblock {\emph{\JournalTitle{Phys. Rev. Mater.}}}
  \textbf{\bibinfo{volume}{6}}, \bibinfo{pages}{044403},
  \doiprefix\url{10.1103/PhysRevMaterials.6.044403} (\bibinfo{year}{2022}).

\bibitem{Wakatsuki2017}
\bibinfo{author}{Wakatsuki, R.} \emph{et~al.}
\newblock \bibinfo{journal}{\bibinfo{title}{{Nonreciprocal charge transport in
  noncentrosymmetric superconductors}}}.
\newblock {\emph{\JournalTitle{Science Advances}}}
  \textbf{\bibinfo{volume}{3}}, \doiprefix\url{10.1126/sciadv.1602390}
  (\bibinfo{year}{2017}).

\end{thebibliography}

\end{document}